\documentclass[a4paper]{aa}
\usepackage{natbib}
\usepackage{graphicx}
\usepackage{color}
\usepackage{ulem}

\newcommand{\chui}[1]{#1}
\newcommand{\warn}[1]{#1}

\newcommand{\total}[1]{678}
\newcommand{\ndebris}[1]{53}
\newcommand{\nlarge}[1]{nine}
\newcommand{\nIRSF}[1]{325}
\newcommand{\klimit}[1]{Ks$<$4.5}
\newcommand{\aveimprove}[1]{14\% to 1.8\%}
\newcommand{\newobj}[1]{eight}
\newcommand{\conf}[1]{28}
\newcommand{\prevdebris}[1]{17}

\title{Faint warm debris disks around nearby bright stars 
explored by AKARI and IRSF}
\author{
Daisuke Ishihara$^{1}$,
Nami Takeuchi$^{1}$,
Hiroshi Kobayashi$^{1}$,
Takahiro Nagayama$^{1,2}$,\\
Hidehiro Kaneda$^{1}$,
Shu-ichiro Inutsuka$^{1}$,
Hideaki Fujiwara$^{3}$,
and
Takashi Onaka$^{4}$
}
\institute{ 
Department of Physics,
Nagoya University, Furo-cho, Chikusa-ku, Nagoya, Aichi, 464-8602, Japan
\and 
Department of Physics and Astronomy,
Kagoshima University, 1-21-35, Korimoto, Kagoshima, 890-0065, Japan
\and 
Subaru Telescope, National Astronomical Observatory of Japan, 
650 North A’ohoku University Park, PA 16802, USA
\and 
Department of Astronomy, Graduate School of Science,
University of Tokyo, 7-3-1 Hongo, Bunkyo-ku, Tokyo, 113-0033, Japan
}

\abstract
{ 
Debris disks are important observational clues 
for understanding planetary-system formation process. 
In particular, faint warm debris disks may be related to late planet
formation near 1 au. 
A systematic search of faint warm debris disks is necessary to reveal
terrestrial planet formation. 
}
{ 
Faint warm debris disks show excess emission that peaks at mid-IR wavelengths.
Thus we explore
debris disks using the AKARI mid-IR all-sky point source catalog (PSC),
a product of the second generation unbiased IR all-sky survey.
}
{ 
We investigate IR excess emission
for \total{} isolated main-sequence stars 
for which there are 18\,$\mu$m detections in the AKARI mid-IR all-sky catalog
by comparing their fluxes with the predicted fluxes
of the photospheres based on
optical to near-IR fluxes and model spectra.
The near-IR fluxes are first taken from the 2MASS PSC.
However, 286 stars with \klimit{} in our sample
have large flux errors in the 2MASS photometry due to saturation.
Thus we have measured accurate J, H, and Ks band fluxes,
applying neutral density (ND) filters for 
Simultaneous InfraRed Imager for Unbiased Survey (SIRIUS) on IRSF, 
the $\phi$1.4\,m near-IR telescope in South Africa,
and improved the flux accuracy from 14\% to 1.8\% on average.
}
{ 
We identified \ndebris{} debris-disk candidates
including eight new detections from our sample of \total{} main-sequence stars.
The detection rate of debris disks for this work is $\sim$8\%,
which is comparable with those in previous works by Spitzer and Herschel.
}
{ 
The importance of this study is the detection of 
faint warm debris disks around nearby field stars.
At least \nlarge{} objects have a large amount of dust for their ages, 
which cannot be explained by the conventional steady-state collisional cascade model.
}

\authorrunning{Ishihara, D. et al.}
\titlerunning{Warm debris disks explored by AKARI and IRSF}

\bibpunct{(}{)}{;}{a}{}{,}

\begin{document}
\maketitle

\section{Introduction} \label{intro}
Debris disks are optically thin circumstellar dust disks
around main-sequence stars whose
proto-planetary disks have been dissipated.
One of the possible scenarios is that
they provide important observational clues for understanding 
planetary-system formation process
because the dust grains in these systems may be supplied
by collisions between planetesimals or 
growing proto-planets \citep[e.g.,][]{Wyatt08}. 

Observationally, debris disks are detected
as infrared (IR) excess 
emission above the expected photospheric emission
of dwarf stars.
Debris disks were first reported in 1980s
based on the Infrared Astronomical Satellite (IRAS) infrared all-sky survey observations.
Including Vega \citep{Aumann},
an appreciable number of IR excess stars were reported \citep{Oudmaijer, Mannings}.
Most of them have excess emission at 
a wavelength of 60\,$\mu$m.
Subsequently, in the 1990s, the detection rate and 
the timescale of dissipation of debris disks 
were investigated \citep{Habing, Spangler}
using the Infrared Space Observatory (ISO).
In the last decade, the Spitzer and Herschel space telescopes 
detected hundreds of debris disks.
%
The detection rate of debris disks is found to be 
10--30\% \citep[e.g.,][]{Trilling,Eiroa}.

The time scale for the dissipation of debris disks,
which reflects the time scale of planetary-system formation,
is statistically investigated for A-type and FGK-type stars, separately
\citep[e.g.,][]{Rieke,Su,Siegler}.
Even in the era of observatory-type IR telescopes,
the importance of unbiased surveys is realized 
through the studies based on the IRAS database \citep[e.g.,][]{Rhee}.
In 2010s, the {\it Wide-field Infrared Survey Explorer} \citep[WISE;][]{Wright10}
has explored large amount of debris disks with a highly sensitive all-sky survey
\citep[e.g.,][]{Patel}.

Though the number of stars in debris-disk samples has been increased,
mysteries still remain over in essential points.
%
For example, no simple statistical relation is found 
between debris disks and planet-hosting stars
\citep{Greaves,Moro-Martin,Kospal,Bryden09,Dodson,Moro-Martin15},
though debris disks must be observational clues to ongoing planetary-system formation.
This might be because the current exoplanet samples and debris-disk samples
are weighted toward hot-Jupiters in close orbits 
and younger, heavier disks, respectively.
In addition to the Asteroid and Kuiper belts, our solar system of 4.6\,Gyrs old 
also has an optically thin dust disk named the Zodiacal cloud
\citep[e.g.,][]{Kelsall,BMay,Planck}.
The relation between the Zodiacal cloud and debris disks is also 
an important subject for discussion.
One of the promising approaches for these subjects
is a systematic exploration of faint debris disks in inner orbits
at late stages of planetary system formation.

AKARI is the first Japanese IR astronomical satellite \citep{Murakami}
with a 70\,cm-diameter 6\,K telescope \citep{Kaneda}.
The AKARI mid-IR all-sky survey was performed
with two photometric bands centered 
at wavelengths of 9 and 18\,$\mu$m
using one of the on-board instruments, 
the Infrared Camera \citep[IRC;][]{Onaka}
simultaneously with the far-IR survey
conducted at the 65, 90, 140, and 160\,$\mu$m bands 
\citep{Kawada}. 
The AKARI all-sky survey is the second generation 
unbiased all-sky observation in the IR following the IRAS survey.
The publicly available mid-IR all-sky 
point source catalog \citep[PSC;][]{Ishihara10}
contains a large amount of newly detected 
IR sources \citep[e.g.,][]{Ishihara11},
as a result of 
improvements in sensitivity and spatial resolution
over the IR all-sky survey by IRAS \citep{Neugebauer}.
Among the AKARI bands, the 18\,$\mu$m band is sensitive to the
IR radiation from warm dust grains with temperatures of 100--300\,K,
which are comparable to the equilibrium temperatures
for dust grains at around 1\,AU from a solar-type star.
The detection limit for the AKARI 18\,$\mu$m band is 90\,mJy.
In total, 194,551 objects in the PSC have 18\,$\mu$m detections.
In the previous study using the AKARI PSC,
we reported 24 debris-disk candidates
with large excess emission in the AKARI 18\,$\mu$m band 
based on conservative criteria \citep{Fujiwara13}.
Various kinds of minerals were detected through 
by follow-up observations of newly detected debris-disk candidates.
Their conditions for formation give us information
on events in the planetary-system formation stages
\citep{Fujiwara10a,Fujiwara10b}.
Thus, further systematic exploration of debris disks based on this database is warranted.

In this paper, we explore debris-disk candidates 
using the AKARI/IRC mid-IR PSC ver.~1
to enable statistical discussions on the evolution of debris disks,
their relation to planetary system formations, 
and the relation between debris disks and the zodiacal light.

\section{Observations and data analyses} \label{obs}
Debris disks are detected as IR excess emission 
around main-sequence stars. 
First, we list known main-sequence stars with 18\,$\mu$m detections.
Then we predict their photospheric fluxes at 18\,$\mu$m
based on the optical to near-IR fluxes and model spectra.
Finally, we compare the predicted fluxes with the observed fluxes
and investigate excess emission.

\subsection{Sample selection} \label{sample}

We first obtain 1,735 main-sequence candidates 
that have AKARI 18\,$\mu$m fluxes.
\chui{977} objects are selected from the Tycho-2 spectral type catalog \citep{Wright} while
\chui{758} objects are from the Hipparcos catalog \citep{Perryman}.
We select B8V--M9V stars
based on the Tycho-2 spectral type catalog \citep{Wright},
which contains the largest number of stars 
with information on the luminosity class.
The Tycho-2 spectral type catalog is made by combining other original catalogs.
The luminosity classes for most of the stars are
quoted from Michigan catalog for HD stars, Vol.~1--5 \citep{Houk1,Houk2,Houk3,Houk4,Houk5}.
These catalogs cover the southern hemisphere (Dec. $<+5^\circ$)
with the limiting magnitude of V$\sim$15\,mag.,
which is deep enough to cover 
all the main-sequence stars detected by the AKARI mid-IR survey.
To cover stars in the northern hemisphere,
we also search for main-sequence stars from the Hertzsprung-Russell (HR) diagram
made by using the Hipparcos catalog \citep{Perryman}.
Stars located in the main-sequence locus
($M_V < 6.0\times(B-V) - 2.0$) are added to our sample.
%
%
A total 64,209 main-sequence candidates are listed
and cross-identified with the AKARI mid-IR PSC sources
using a search radius of 3$''$,
because the astrometric accuracy for 
the Hipparcos catalog, Tycho-2 spectral catalog, and the AKARI mid-IR PSC
are $\sim$0.7\,milli-arcsecond, $\sim$0.5\,$''$, and 2$''$, respectively
\citep{Perryman,Wright,Ishihara10}.
%

Then we carefully screen the 1,735 targets using the SIMBAD database 
to make a clean sample of isolated main-sequence stars. 
Table~\ref{tbl:simbad} summarizes the classification
of our main-sequence candidates by the SIMBAD database.
Based on the SIMBAD classification,
suspected proto-planetary disks and mass-losing stars are removed from our sample,
because their IR excess emission tends to be misinterpreted as signs of debris disks.
Suspected binary stars, multiple stars, and stars in clusters 
are also rejected from our sample 
because it requires further detailed analyses 
to evaluate contamination in IR fluxes 
from their companion or neighboring objects. 
Finally, \chui{750} objects 
classified as star, high-proper motion star, and extra-solar planet candidate
are selected for our analyses.


\begin{table}[!h]
\caption{Classification of our dwarf sample by the SIMBAD database.}
\label{tbl:simbad}
\begin{center}
\begin{tabular}{clr}\hline\hline
\multicolumn{2}{l}{Category}       & Num. of stars \\
\hline
\multicolumn{3}{l}{Isolated main-sequence stars}   \\
\ & Star ($\ast$)                            & 476 \\
\ & High proper-motion star (PM$\ast$)       & 254 \\
\ & Extra-solar planet candidate (Pl$?$)     &  20 \\
\multicolumn{3}{l}{Stars in a multiple system or a cluster}\\
\ & Spectroscopic binary (SB$\ast$)          & 170 \\
\ & Double or multiple star ($\ast\ast$)     & 130 \\
\ & Star in a multiple system ($\ast$i$\ast$)& 181 \\
\ & Star in a cluster ($\ast$iC)             &  18 \\
\ & Star in a nebula ($\ast$iN)              &   5 \\
\multicolumn{2}{l}{Variable stars $^\dagger$}      &  332 \\
\multicolumn{2}{l}{YSOs $^\ddagger$}           &  141 \\
\multicolumn{2}{l}{Old stars $^\mathsection$}     &    6 \\
\multicolumn{2}{l}{Others (IR, Rad)}         &    2 \\
\hline
\multicolumn{2}{l}{Total}                   & 1,735 \\
\hline
\end{tabular}
\end{center}
$\dagger$       ... Including Variable Star (V*), 
Star suspected of Variability (V*?), Variable of BY Dra type (BY*),
Variable of RS CVn type (RS*), Variable Star of delta Sct type (dS*),
Variable Star of gamma Dor type (gD*), Semi-regular pulsating Star (sr*),
Variable Star of alpha2 CVn type (a2*), Pulsating variable Star (Pu*),
Ellipsoidal variable Star (El*), Variable Star of RV Tau type (RV*),
Variable Star with rapid variations (RI*), Rotationally variable Star (Ro*),
Variable Star of beta Cep type (bC*), Variable Star of Mira Cet type (Mi*),
Eclipsing binary of W UMa type (contact binary) (WU*), Variable Star of W Vir type (WV*),
Variable Star of R CrB type (RC*), Eruptive variable Star (Er*),
Variable Star of Orion Type (Or*), Cepheid variable Star (Ce*),
Eclipsing binary of Algol type (detached) (Al*), Eclipsing binary 
of beta Lyr type (semi-detached) (bL*).\\
$\ddagger$ ... Contains T Tau-type star (TT$\ast$),
  pre-main sequence star (pr$\ast$), herbig-haro object (HH),
  flare star (Fl$\ast$), emission-line star (Em$\ast$),
   Be star (Be$\ast$).\\
$\mathsection$ ... Contains carbon star (C$\ast$), 
  planetary nebula (PN), Wolf-Rayet (WR$\ast$), 
  post-AGB star (pA$\ast$), white dwarf (WD$\ast$).
\end{table}

\subsection{Photometric data for central stars} \label{phot}
We create optical to near-IR spectral energy distributions (SEDs) 
of the central stars 
using archival data. 
The SED for each central star contains five to seven photometric fluxes.
The B$_{\rm T}$  and V$_{\rm T}$ band fluxes are 
taken from the Tycho-2 spectral type catalog
\citep{Wright}. 
The J, H, and Ks band fluxes are taken from 2MASS PSC ver.~6 \citep{Cutri}.
We also add the R-, and I-band fluxes from 
the Catalog of stellar photometry in Johnson's 11-color system \citep{Ducati}, if available. 

\subsection{J, H, Ks photometry by IRSF/SIRIUS} \label{irsf}

\begin{figure*}
\includegraphics[width=6cm]{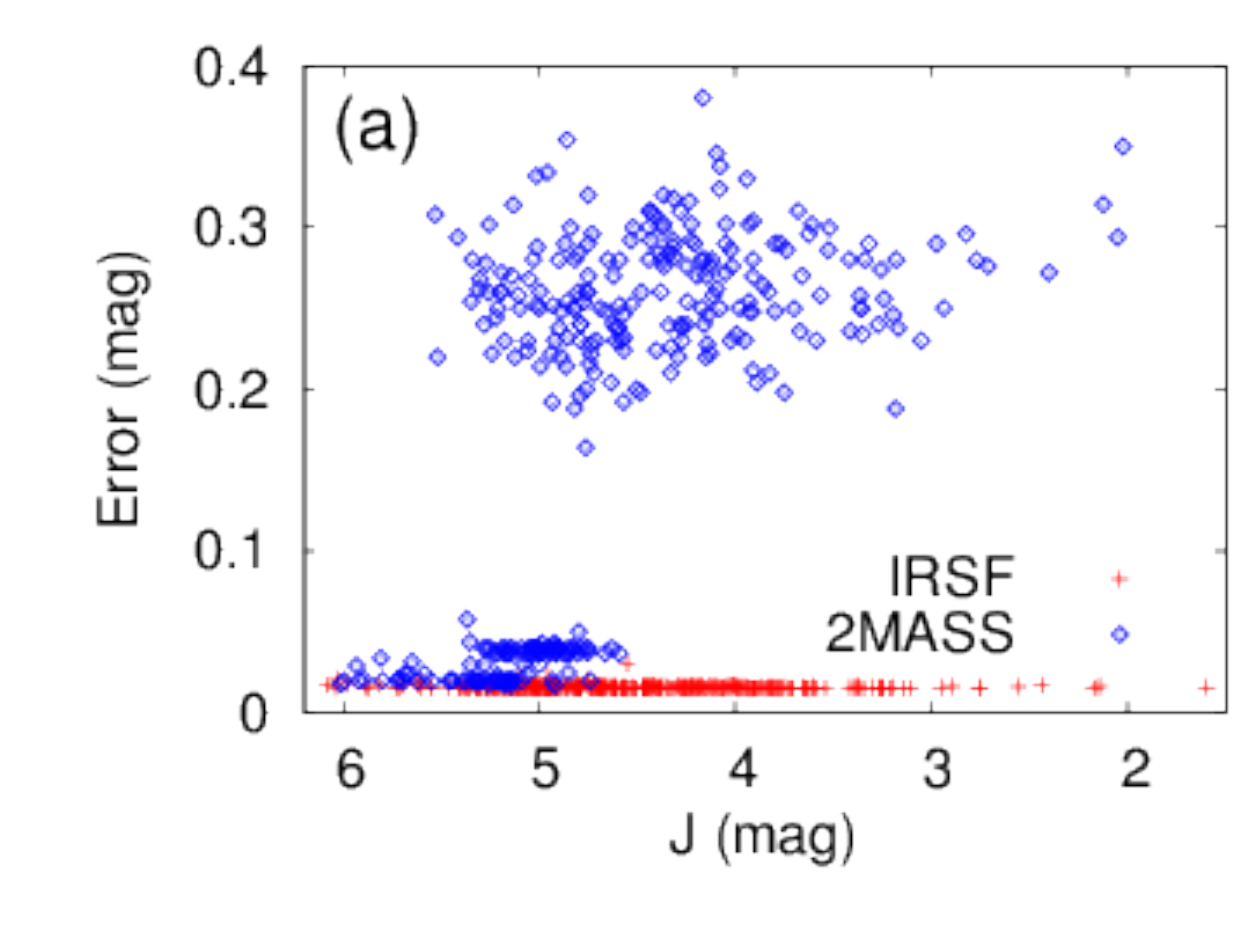}
\includegraphics[width=6cm]{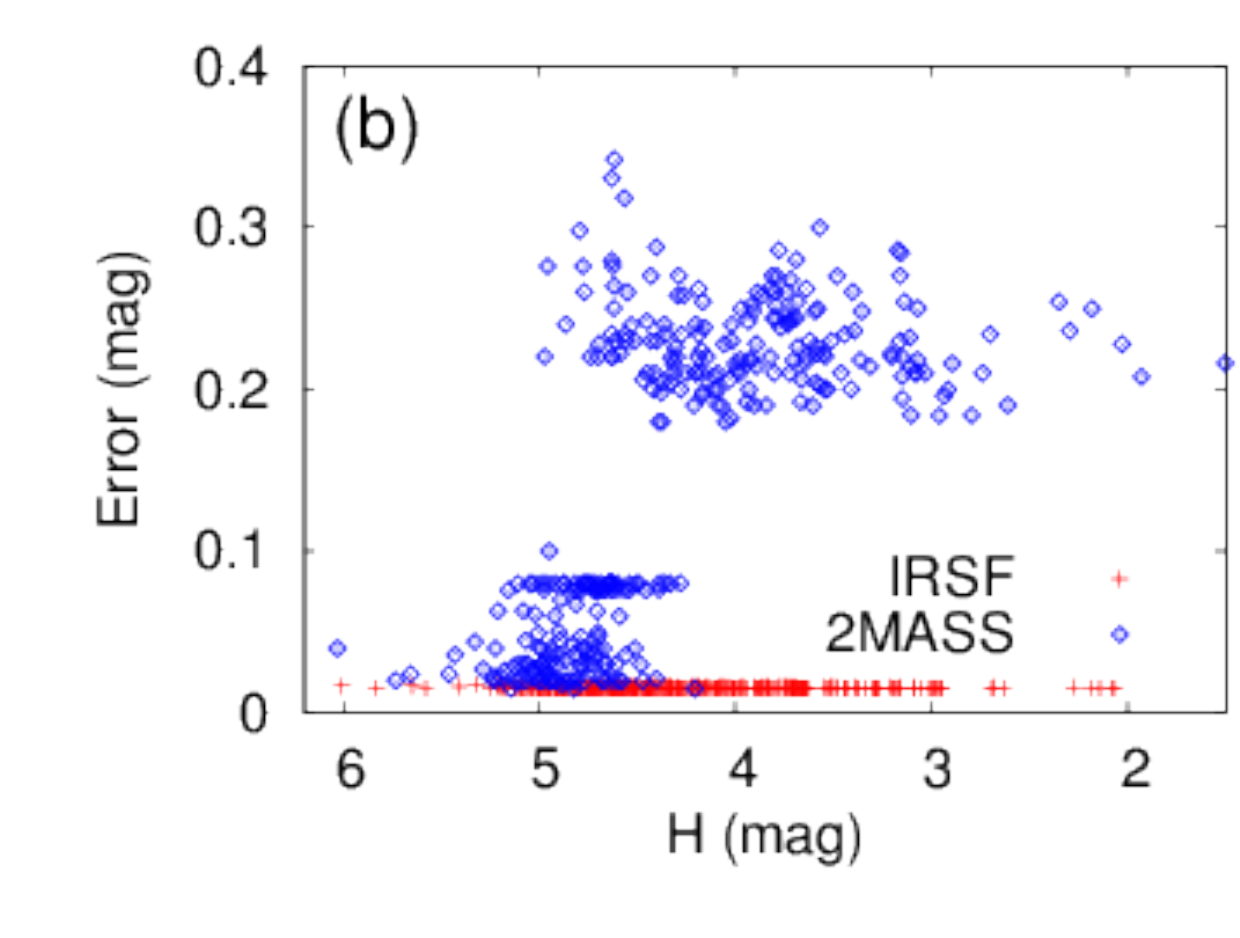}
\includegraphics[width=6cm]{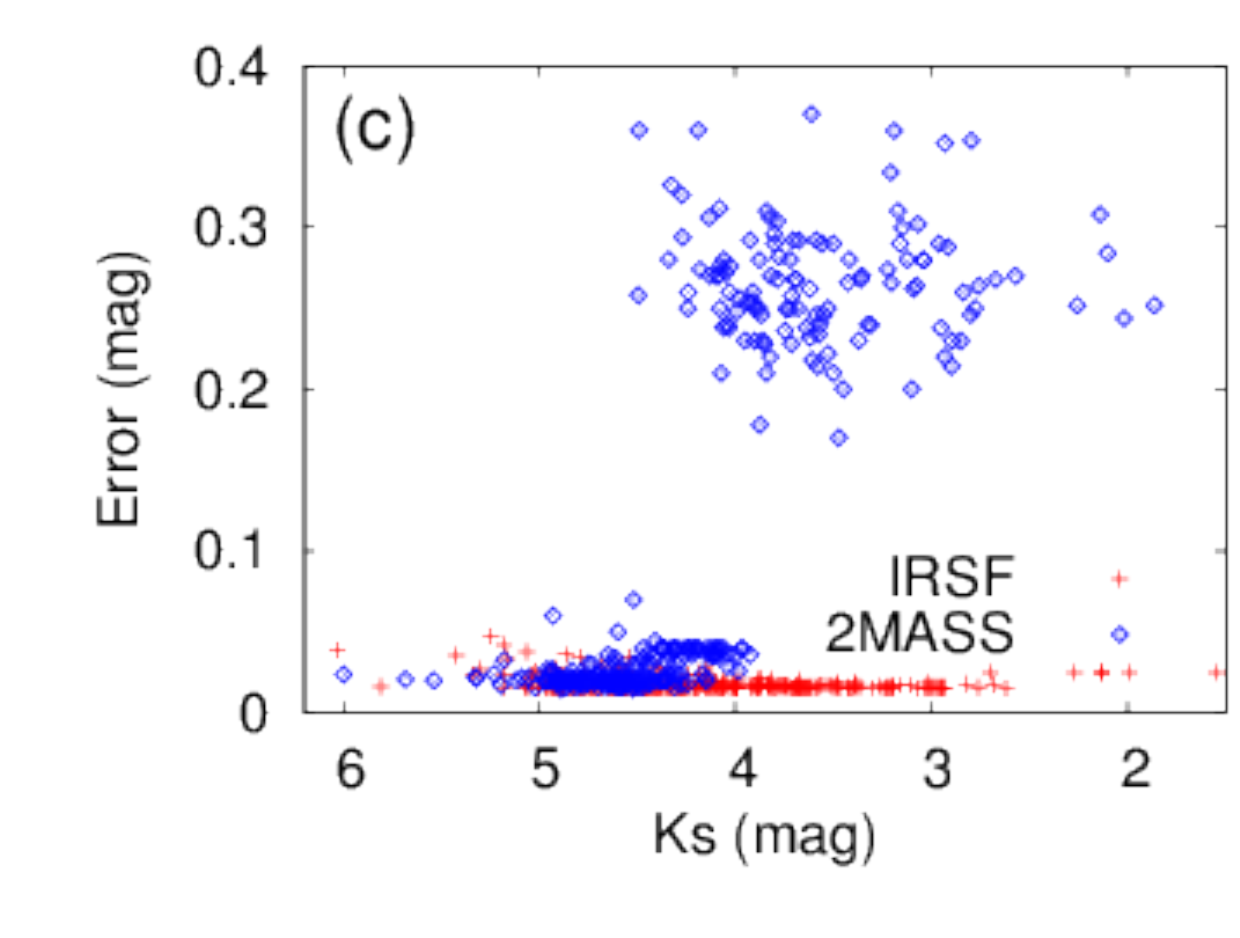}
\caption{
Measurement errors from 2MASS (squares) and IRSF (crosses)
as a function of the magnitude in the 
(a) J band, (b) H band, and (c) Ks band.
}\label{fig:err}
\end{figure*}

Detection of IR excess emission needs
accurate estimation of photospheric emission 
as well as accurate measurement in the mid-IR.
The fluxes in the J, H, and Ks bands play
important roles to determine photospheric emission accurately.
Stars in our sample have fluxes of magnitude one to six in the Ks band.
Most of them are too bright to measure their fluxes accurately
due to the low saturation limit of 2MASS.
Since bright stars are evaluated by point spread function fitting
of saturated images \citep{Skrutskie},
the measurement errors by 2MASS are as large as 11--17\% in
the J, H, and Ks bands for stars with J$<$5.5, H$<$5.2, or Ks$<$4.5 (see Fig.~\ref{fig:err}).
%
%
Thus we have improved the accuracy of photometry
of \nIRSF{} bright main-sequence stars in the J, H, and Ks bands
%
by using the 
Simultaneous InfraRed Imager for Unbiased Survey (SIRIUS) on 
InfraRed Survey Facility (IRSF), where
SIRIUS is a wide field ($7'\times7'$) near-IR camera
which enables simultaneous observations in the J, H, and Ks bands 
\citep{Nagayama}
and
IRSF is the $\phi$ 1.4\,m near-IR telescope 
located at Sutherland in South Africa and 
managed by Nagoya University \citep{Sato}. 
%
Table~\ref{tbl:jhk} summarizes photometric results for our sample observed by IRSF.
With these observations we have successfully improved
the J, H, and Ks flux errors
from 17\% to 1.9\%,  14\% to 1.4\%, and  11\% to 2.0\%, respectively
(see Fig.~\ref{fig:err}).
The details of the observations are given in Appendix~\ref{A1}.

\subsection{Removing suspected giant stars based on near-IR colors} \label{giants}

By using the improved J, H, and Ks fluxes,
\chui{39} suspected late-type giant stars are removed from our sample
because they show giant-like colors 
in the J--H versus H--Ks color-color diagrams (see Fig.~\ref{fig:JHK}b).
Giants with spectral type later than K1 show different JHK colors from dwarfs
while earlier-type giants show similar JHK colors as dwarfs in the JHK color-color plane. 
We also remove similarly \chui{33} suspected giant stars from the 2MASS based sample (see Fig.~\ref{fig:JHK}a).
Finally, a main-sequence sample with \total{} stars is obtained.

\begin{table}
\caption{J, H, Ks magnitudes of bright main-sequence stars measured by IRSF.}
\label{tbl:jhk} 
\begin{tabular}{lccc}
\hline\hline
IRC name & J (mag.) & H (mag.) & Ks (mag.)\\
\hline
0002575-200245 & 5.34$\pm$0.02 & 5.11$\pm$0.02 & 5.07$\pm$0.02\\
0003444-172009 & 4.62$\pm$0.02 & 4.60$\pm$0.02 & 4.60$\pm$0.02\\
0006195-490431 & 4.78$\pm$0.02 & 4.52$\pm$0.02 & 4.47$\pm$0.02\\
0011158-152806 & 3.98$\pm$0.02 & 3.69$\pm$0.02 & 3.67$\pm$0.02\\
0011441-350758 & 4.41$\pm$0.02 & 4.19$\pm$0.02 & 4.16$\pm$0.02\\
0016140-795104 & 5.32$\pm$0.02 & 4.95$\pm$0.02 & 4.88$\pm$0.02\\
0026122-434043 & 3.65$\pm$0.02 & 3.55$\pm$0.02 & 3.52$\pm$0.02\\
... & & &\\\hline
\end{tabular}
\tablefoot{Table~\ref{tbl:jhk} is available at the CDS.}
\end{table}

\subsection{Estimation of 18\,$\mu$m photospheric flux}
The photospheric flux densities of our sample
are calculated from a Kurucz model \citep{Kurucz}
fitted to the optical to near-IR photometry of the stars
taking extinction into account.
In the fitting, the scale factor $S$ 
(i.e., the distance indicator related to the distance between the star and the Earth), 
and the visual extinction $A_{\rm V}$ are set as a free parameter,
while the effective temperature $T_{\rm eff}$, metallicity $M$, and $\log(g)$ 
are selected 
from discrete sets of template\footnote{ftp://ftp.stsci.edu/cdbs/grid/k93models}
values ($g$ is surface gravity in cgs units).
The error caused by using quantized parameters for the fitting is discussed in Appendix~\ref{A2b}.
The quantized parameters cause uncertainties for individual $F_*$,
but they are much smaller than the systematic uncertainties
for the excess identification.
The range of $A_{\rm V}$ is limited between 0 and 0.5 
under the assumption 
that $A_{\rm V}$ is small in most directions within D$<$100\,pc \citep{Lalement}.
$T_{\rm eff}$ is allowed to vary within $\pm$1,000\,K around the initial value
that is quoted from the SIMBAD database or estimated from the spectral type.
$\log(g)$ is selected among 4,0, 4.5, and 5.0,
because $\log(g)$ for dwarfs varies from 4.2 to 4.7 \citep{AQ}.
We assume solar metallicity for all the objects
because they are located within 600~pc and no significant change 
in metallicity is expected.
For the foreground extinction,
we adopt the extinction curve of
\begin{equation}
\frac{A_\lambda}{A_V} = \frac{0.349+2.087R_V}{1+(\lambda/\lambda_0)^\alpha}\cdot\frac{1}{R_V},
\end{equation}
where $\lambda_0=0.507$\,$\mu$m (i.e., V band),
$A_\lambda$ and $A_V$ are the extinction at $\lambda$ and in the V band,
respectively, and $R_V\equiv A_V / E(B-V)$.
This is the generalized extinction curve given by \citet{Fitz} based 
on the model by \citet{Pei}. 
%
%
The dust properties in the lines of sight determine $R_V$ and $\alpha$.
We use $\alpha=2.05$ and $R_V=3.11$ following \citet{Fujiwara13},
assuming that the extinction curve is uniform within our survey volume.
Though this extinction curve does not take into consideration the silicate feature at 10~$\mu$m,
it does not affect the SED fitting of the photosphere
because we only use 0.3--2.4~$\mu$m (from U to Ks bands) for the fitting.
We might overestimate the photospheric emission at 18~$\mu$m
because the model \citep{Fitz} does not take account of the silicate 18~$\mu$m feature.
It works as a conservative estimate for the IR excess identification.
\chui{The reliability of the fitting results is discussed in Appendix~\ref{A2}.}

\subsection{AKARI 18\,$\mu$m photometry}

At the next step,
we compare the predicted photospheric fluxes ($F_{\rm 18,*}$) 
with the fluxes ($F_{\rm 18,obs}$) observed at $\lambda=18$\,$\mu$m
using the AKARI mid-IR PSC.
%
%
The monochromatic fluxes in the PSC are derived for objects
with spectra of $F_\lambda\propto\lambda^{-1}$.
We apply color corrections to the catalog values
assuming that the spectra of the photospheres of main-sequence stars
are give by $F_\lambda\propto\lambda^{-4}$.

%
%

\subsection{Excess identification} \label{sec:excess}
\begin{figure*}
\begin{center}
\includegraphics[width=9cm]{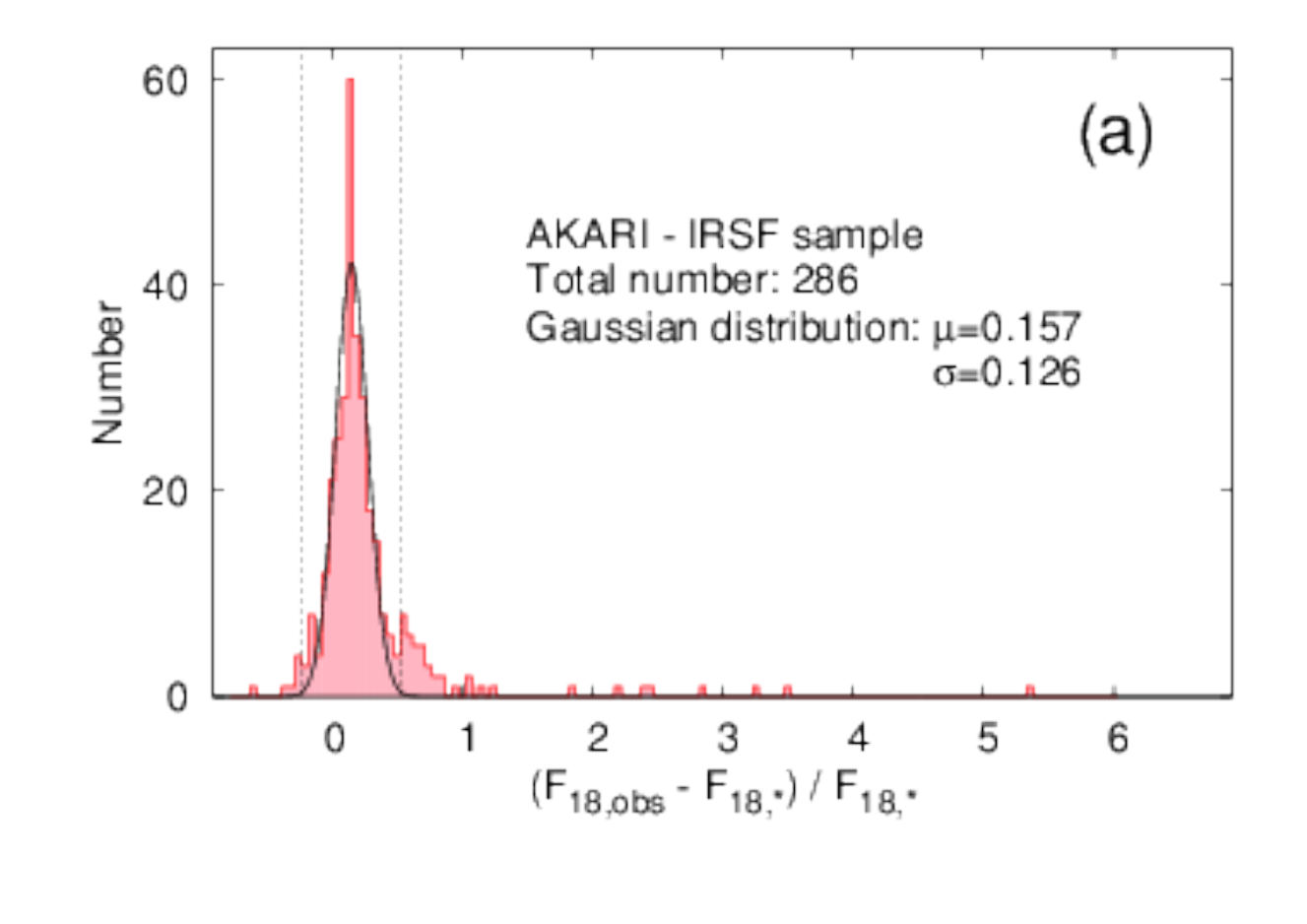}
\includegraphics[width=9cm]{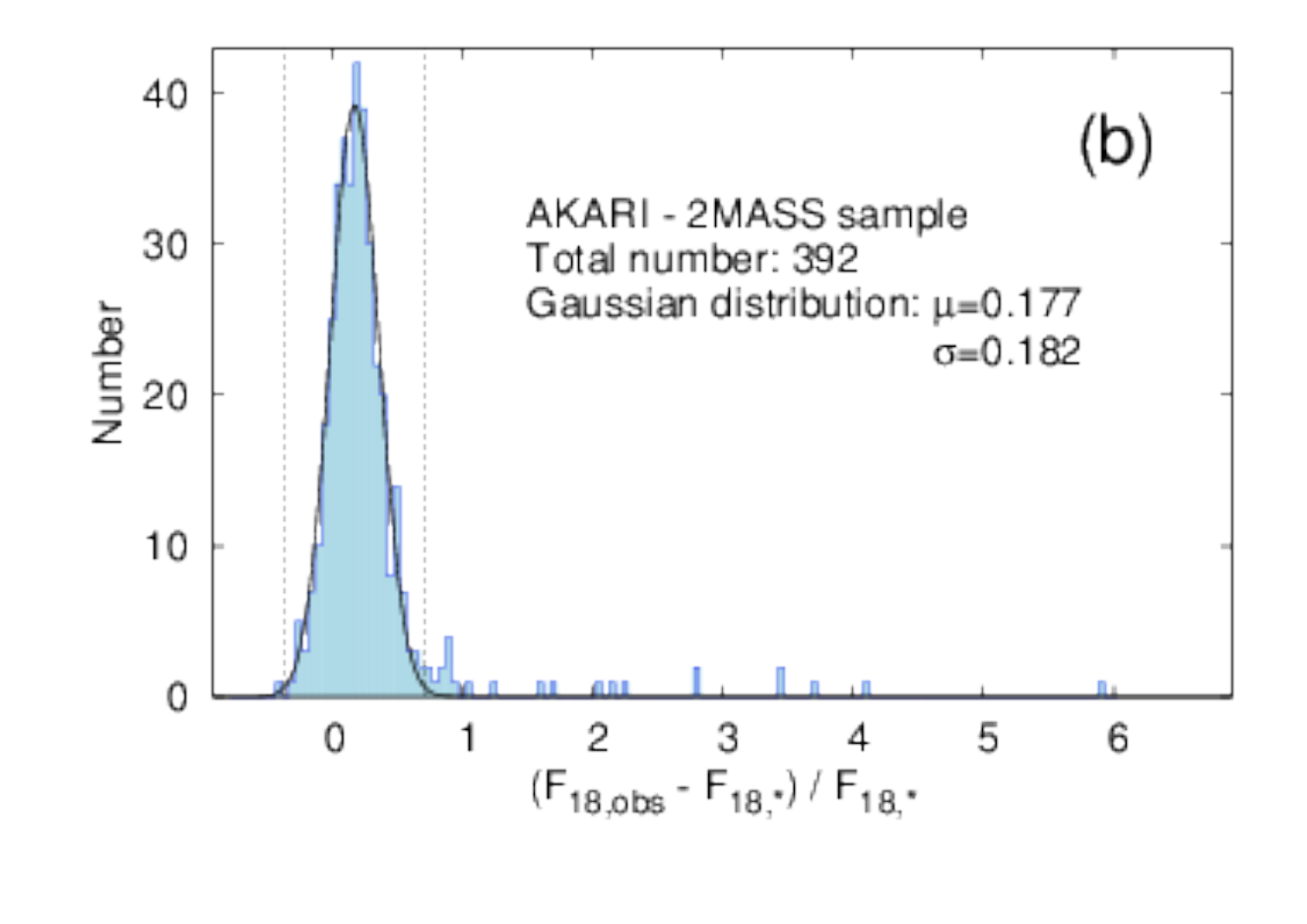}
\caption{(a) Distributions of 
the excess ratio ($(F_{\rm 18,obs}-F_{\rm 18,\ast})/F_{\rm 18,\ast}$)
for our sample with IRSF measurements.
The center of the peak, $\mu$, and the standard deviation, $\sigma$, 
of the Gaussian distribution are indicated in the graphs.
(b) Same as (a), but for the sample with 2MASS measurements.
}\label{fig:judge}
\end{center}
\end{figure*}

We investigate 18\,$\mu$m excess emission for each star as follows:
The excess ratio at 18\,$\mu$m,
$(F_{\rm 18,obs}-F_{\rm 18,*})/F_{\rm 18,*}$, is calculated for each star.
%
Then we make histograms of the excess ratios
for the AKARI--IRSF sample and the AKARI--2MASS sample, separately,
as shown in Fig.~\ref{fig:judge}. 
We fit these histograms with a Gaussian,
assuming that these distributions are mainly caused by photon noise.
We obtain the center of the peak
$\mu = 0.157\pm0.005$ and 
standard deviation $\sigma = 0.126\pm0.005$ 
of the Gaussian function for the AKARI--IRSF sample,
while
$\mu = 0.177\pm0.003$ and $\sigma = 0.182\pm0.003$ for the AKARI--2MASS sample.
We regard that
$\mu$ and $\sigma$ are
the systematic offset and the total uncertainties of the sample, respectively.
%
We then select objects as debris-disk candidates 
which show excess ratios larger than $\mu+3\sigma$ for both samples.
%

Then, we check the known extragalactic sources
around the excess objects using the NED database
in order to avoid incorrect excess identifications
by chance alignment of background sources.
Even the nearest NED source
(1RXS\,J194816.6+592519 for HD\,187748) aparts as far as 10.02$''$.
For all the other objects, there are no counter parts within 12$''$ 
which corresponds to the twice of the FWHM of the AKARI 18\,$\mu$m PSF 
\citep[5.7$''$;][]{Onaka}).
Finally, for all the debris-disk candidates, we check
the 2MASS K$_s$, AKARI 9\,$\mu$m, and AKARI 18\,$\mu$m images
to investigate the effects of image artifacts,
and the contamination of background or foreground sources.
Details are in Appendix.~\ref{imageana}.

The differences from our previous work \citep{Fujiwara13} 
in the process for identifying debris disks are as follows:
\begin{itemize}
\item Flux accuracy of the central stars:
We have improved flux accuracy of the photosphere
for nearby bright stars (\nIRSF{} objects with $Ks<4.5$)
by follow-up observations using IRSF
instead of using publicly-available 2MASS fluxes.
\item Sample selection:
In this work, we exclude double stars, multiple stars, spectroscopic binaries,
and stars in a cluster as well as suspected YSOs and mass-losing stars
before investigating the excess emission to discuss the excess probability
more accurately.
13 objects out of the 24 debris-disk candidates reported in \citet{Fujiwara13}
are again listed in the current list while
11 objects are not included in our list because they are multiple stars.
\end{itemize}

\section{Results} \label{results}

\subsection{Debris-disk candidates with AKARI 18\,$\mu$m excess emission}
\label{list}

As a result, \ndebris{} objects out of \total{} main-sequence stars
in our sample are identified as debris-disk candidates 
that have excess emission at the AKARI 18\,$\mu$m band.
Tables~\ref{tbl:list} and \ref{tbl:list2} summarize 
the parameters of our debris-disk candidates,
for the 2MASS based sample and IRSF based sample, respectively.
Figure~\ref{fig:SEDs} shows the SEDs of the individual objects.

It should be noted that some objects 
show flux ratios much larger than expected for main sequence stars.
%
HD\,93942, HD\,145263, HD\,165014, HD\,166191, and HD\,167905
are classified as main-sequence stars in literature \citep{Wright}, and
HD\,9186 and HD\,215592 has no luminosity class information in literature
and were identified as main-sequence stars from the location on the HR diagram.
All of them are reported as debris disks candidates 
in previous works \citep{Oudmaijer,Clarke,McDonald,Fujiwara13}.
It should be noted that
HD\,166191 was studied by both
\citet{Schneider} and \citet{Kennedy14}
and the conclusions were different.
Additional observations 
are certainly needed to clarify the nature of these objects.
HD~93942 shows a transitional-disk like SED
composed of photosphere and thick circumstellar emission.
It should also be noted that B- and A-type stars,
HD\,161840, HD\,32509, HD\,9186, HD\,118978, and HD\,28375,
show ambient circumstellar emission on the AKARI 18\,$\mu$m images.
%
%
These objects are marked in Tables 3 and 4.

\begin{table*}[!t]
\caption{List of debris-disk candidates with 
the AKARI 18\,$\mu$m excess emission (the 2MASS based sample).}
\label{tbl:list}
\centering
\begin{tabular}{lcccccccc}\hline\hline
Star name &
IRC name &
\multicolumn{1}{c}{Spectral}&
\multicolumn{1}{c}{Distance}&
Age&
$F_{\rm 18,obs}$$^\$$& 
$F_{\rm 18,*}$$^\$$& 
Excess$^\$$&
References\\
&&type&\multicolumn{1}{c}{pc}&
\multicolumn{1}{c}{Gyr}&
\multicolumn{1}{c}{Jy}&
\multicolumn{1}{c}{Jy}&ratio&\\
\multicolumn{1}{c}{(1)}&(2)&(3)&(4)&(5)&(6)&
\multicolumn{1}{c}{(7)}&(8)&(9)\\ 
\hline
HD\,9186   & 0132297+675740 &  (B9)      & 300$\pm$80  &  --        & 0.341 & 0.00775 & 43$^*$$^\ddagger$    & g \\
HD\,9672   & 0134378-154034 &  A1V       &  61$\pm$3   &  --        & 0.197 & 0.0873  & 1.26  & a,g,h \\
HD\,16485  & 0240209+493337 &  B9V       & 500$\pm$200 &  --        & 0.191 & 0.0481  & 2.97  & g \\
HD\,26912  & 0415320+085332 & B3V (B3IV) & 130$\pm$20  &  --        & 0.351 & 0.205   & 0.708 & a,g \\
HD\,28375  & 0428321+012250 &  (B3V)     & 120$\pm$20  &  --        & 0.570 & 0.0586  & 8.73$^\ddagger$  & a,g \\
HD\,32509  & 0504500+264315 &  (A2e)     & 150$\pm$30  &  --        & 0.256 & 0.0223  & 10.5$^\ddagger$  & a,g\\ 
HD\,34890  & 0520289-054843 &  (A0)      & 180$\pm$50  &  --        & 0.166 & 0.00638 & 25    & g\\
HD\,36546  & 0533307+243743 &  (B8)      & 100$\pm$10  &  --        & 0.684 & 0.0252  & 26.1  & g\\
HD\,39415  & 0554415+443007 &  F5V       &  --         &  --        & 0.271 & 0.0197  & 12.7  & a,h\\
HD\,44892  & 0623427-162801 & (A9/F0IV)  & 160$\pm$20  &  --        & 0.580 & 0.0742  & 6.82  & a,g,h\\
HD\,64145  & 0753297+264556 &  A3V       &  78$\pm$6   &  --        & 0.300 & 0.183   & 0.641 & h\\
HD\,65372  & 0758303+025609 &  (A3)      & 240$\pm$50  &  --        & 0.743 & 0.0452  & 15.4  & a,g\\
%
HD\,75416  & 0841193-785747 &  B8V$^\dagger$       &  97$\pm$4   &  --        & 0.165 & 0.0604  & 1.73  & f,g\\
HD\,93942  & 1049236-584703 &  A1V       &  --         &  --        & 34.4  & 0.00279 & 12300$^*$ & b\\ 
HD\,102323 & 1146229-562242 &  (A2IV)    & 400$\pm$200 &  --        & 0.378 & 0.00646 & 57.6$^*$  & a,g\\
HD\,105209 & 1206526-593529 &  A1V       &  --         &  --        & 0.234 & 0.0212  & 10.1  & a,b,h\\
HD\,118978 & 1342010-584712 &  (B9III)   & 200$\pm$30  &  --        & 0.339 & 0.0898  & 2.78$^\ddagger$  & a,g\\
HD\,120780 & 1352354-505518 & K3V (K0V)  &16.4$\pm$0.3 &  --        & 0.199 & 0.123   & 0.61  & \\ 
HD\,121617 & 1357411-470035 &  A1V       &  --         &  --        & 0.321 & 0.0185  & 16.4  & h\\
HD\,145263 & 1610551-253122 &  F0V       & 120$\pm$20  &  --        & 0.481 & 0.0106  & 44.6$^*$  & g,h \\
HD\,146055 & 1615080-243518 &  B9V $^\dagger$       &  --         &  --        & 0.172 & 0.0196  & 7.78  & \\ 
HD\,155401 & 1712251-274544 &  B9V(N)    & 170$\pm$20  &  --        & 0.412 & 0.0429  & 8.61  & a,g\\
HD\,165014 & 1804432-205643 &  F2V       &  --         &  --        & 0.948 & 0.0269  & 34.3$^*$  & b,h\\
HD\,166191 & 1810303-233401 & F3/5V (F4V)& 140$\pm$30  &  --        & 2.44  & 0.0236  & 102$^*$   & a,b,h\\
HD\,167905 & 1818182-232819 &  F3V       &  --         &  --        & 1.75  & 0.0317  & 54.2$^*$  & a,b,h\\
HD\,169666 & 1819080+713104 &  (F5)      &  51$\pm$1   &  2.0$\pm$0.7$^{(1)}$, 2.0$^{(3)}$  & 0.128 & 0.0801 & 0.604 & g\\
HD\,187748 & 1948154+592523 &  (G0)      &28.4$\pm$0.4 & 3.2$^{(3)}$ & 0.197 & 0.115   & 0.718 & \\
HD\,215592 & 2245380+415258 &  (A0)      & 600$\pm$300 &  --        & 0.332 & 0.00953 & 33.9$^*$  & g\\
HD\,222173 & 2338082+431604 &  B8V$^\dagger$       & 150$\pm$20  &  --        & 0.256 & 0.159 & 0.611   & g\\
HD\,225132 & 0003444-172009 &  (B9IVn)   &  70$\pm$4   &  --        & 0.303 & 0.182 & 0.662   & g\\
HD\,279128 & 0352162+332422 &  (B8) $^\dagger$      & 300$\pm$100 &  --        & 0.362 & 0.0232  & 14.6  & g \\
\hline
\end{tabular}
\tablefoot{
Column
(1): HD name.
(2): Source ID in the AKARI mid-IR PSC.
(3): Spectral type quoted from the Tycho-2 spectral type catalog \citep{Wright}.
The definition in the SIMBAD database is written in parenthesis
if it is different from that in the Tycho-2 spectral type catalog.
(4): Distance (pc) converted from the parallax in the Hipparcos catalog \citep{Perryman}.
(5): Stellar age quoted from the literature.
References are,
 (1) \citet{Chen} is based on the lithium abundances,
 (2) \citet{Feltzing}, and (3) \citet{Holmberg}.
(6): AKARI 18\,$\mu$m flux.
(7): Predicted flux of photosphere at 18\,$\mu$m.
(8): Excess ratio calculated as $(F_{\rm 18,obs} - F_{\rm 18,*})/F_{\rm 18,*}$.
(9): References for the excess detection are,
(a) \citet{Oudmaijer}, (b) \citet{Clarke}, 
(c) \citet{Rieke}, (d) \citet{Bryden06}, 
(e) \citet{Trilling}, (f) \citet{Su}, 
(g) \citet{McDonald}, (h) \citet{Fujiwara13}.
(*) Excess ratio show notably large excess as for debris disks.
($\dagger$) $T_{\rm eff}$ of the central star as a result of photosphere fitting is 
significantly different from that expected from the spectral type.
($\ddagger$) diffuse emission component is recognized in the AKARI 18\,$\mu$m image.
($\$$) The typical error for the flux, estimated photospheric flux, excess ratio are
$\sim$6\%, $\sim$2\%, and $\sim$6\%, respectively.
}
\end{table*}

\begin{table*}[!t]
\caption{List of debris-disk candidates with
AKARI 18\,$\mu$m excess emission (the IRSF based sample).}
\label{tbl:list2}
\centering
\begin{tabular}{lcccccccc}\hline\hline
Star name &
IRC name &
\multicolumn{1}{c}{Spectral}&
\multicolumn{1}{c}{Distance}&
Age&
$F_{\rm 18,obs}$ & 
$F_{\rm 18,*}$ & 
Excess &
References\\
&&type&\multicolumn{1}{c}{pc}&
\multicolumn{1}{c}{Gyr}&
\multicolumn{1}{c}{Jy}&
\multicolumn{1}{c}{Jy}&ratio&\\
\multicolumn{1}{c}{(1)}&(2)&(3)&(4)&(5)&(6)&
\multicolumn{1}{c}{(7)}&(8)&(9)\\
\hline
HD\,1237   & 0016140-795104 &  (G8V)       & 17.5$\pm$0.2 &  11.9$^{(3)}$             & 0.222 & 0.158 & 0.41 & d,e,g\\
HD\,10939  & 0146064-533119 &  A1V         &   57$\pm$2   &  --                      & 0.251 & 0.129 & 0.943 & g\\ 
HD\,39060  & 0547170-510359 & A5V (A6V)    & 19.3$\pm$0.2 &  --                      & 4.76 & 0.559 & 7.5 & a,c,f,g,h\\ 
HD\,50506  & 0640026-804848 &  (A5III)     &  124$\pm$7   &  --                      & 0.209 & 0.133 & 0.575 & g\\ 
HD\,62952  & 0745568-143350 &  F2V         &   72$\pm$4   &  0.8$^{(3)}$              & 0.391 & 0.281 & 0.393 & g\\ 
HD\,69897  & 0820038+271300 &  F6V         & 18.1$\pm$0.3 &  3.3$^{(2)}$, 3.2$^{(3)}$  & 0.524 & 0.357 & 0.469 & \\ 
HD\,89125  & 1017143+230621 &  F8VBW       & 22.7$\pm$0.4 &  6.5$^{(2)}$, 5.5$^{(3)}$  & 0.364 & 0.2 & 0.818 & h\\ 
HD\,99022  & 1123081-564645 &  (A4:p)      &  200$\pm$30  &  --                       & 0.103 & 0.0591 & 0.736 & g\\ 
HD\,101563 & 1141082-291145 &  (G0V)       &   42$\pm$1   &  4.8$^{(3)}$               & 0.203 & 0.135 & 0.503 & \\ 
HD\,106797 & 1217062-654135 &  A0V         &  103$\pm$6   &  --                       & 0.146 & 0.0491 & 1.98 & g,h\\ 
HD\,110058 & 1239461-491156 &  A0V         &  100$\pm$10  &  --                       & 0.0827 & 0.0125 & 5.6 & g,h\\ 
HD\,112060 & 1253320+192850 &  (G5IV)      &   44$\pm$2   &  --                       & 0.307 & 0.193 & 0.59 &  \\
HD\,113457 & 1305023-642630 &  A0V         &   95$\pm$6   &  --                       & 0.107 & 0.0275 & 2.88 & g,h \\ 
HD\,134060 & 1510446-612520 &G2V (G0VFe+04)& 24.1$\pm$0.4 &  9$\pm$3$^{1}$, 7.5$^{(3)}$& 0.239 & 0.161 & 0.485 &  \\
HD\,135379 & 1517307-584805 &  (A3Va)      &   30$\pm$1   &  --                       & 0.536 & 0.376 & 0.423 & g\\ 
HD\,152614 & 1654004+100954 &  (B8V)       &   72$\pm$4   &  --                       & 0.274 & 0.18 & 0.519 & g\\ 
HD\,159492 & 1738054-543002 & (A5IV-V)     &   42$\pm$1   &  --                       & 0.238 & 0.155 & 0.533 & g\\ 
HD\,161840 & 1749105-314211 &  (B8V)$^\dagger$       &  190$\pm$30  &  --                       & 0.384 & 0.157 & 1.45$^\ddagger$ & g \\ 
HD\,172555 & 1845269-645217 &  (A7V)       &   29$\pm$1   &  --                       & 0.912 & 0.25 & 2.64 & a, g\\ 
HD\,176638 & 1903069-420542 &B9/A0V (B9.5V)&   56$\pm$3   &  --                       & 0.225 & 0.156 & 0.44 & g\\ 
HD\,190580 & 2008095-523440 &  G3V         &   58$\pm$3   &  2.8$\pm$0.4$^{(1)}$, 2.9$^{(3)}$  & 0.232 & 0.165 & 0.412 & g\\ 
HD\,193307 & 2021406-495959 &  G0V         &   32$\pm$1   &  9.1$\pm$0.8$^{(1)}$, 7.9$^{(3)}$  & 0.203 & 0.15 & 0.351 & \\ 
\hline
\end{tabular}
\tablefoot{Same as Table~\ref{tbl:list}.}
\end{table*}

\begin{figure*}[!t]
\centering
\includegraphics[width=17cm]{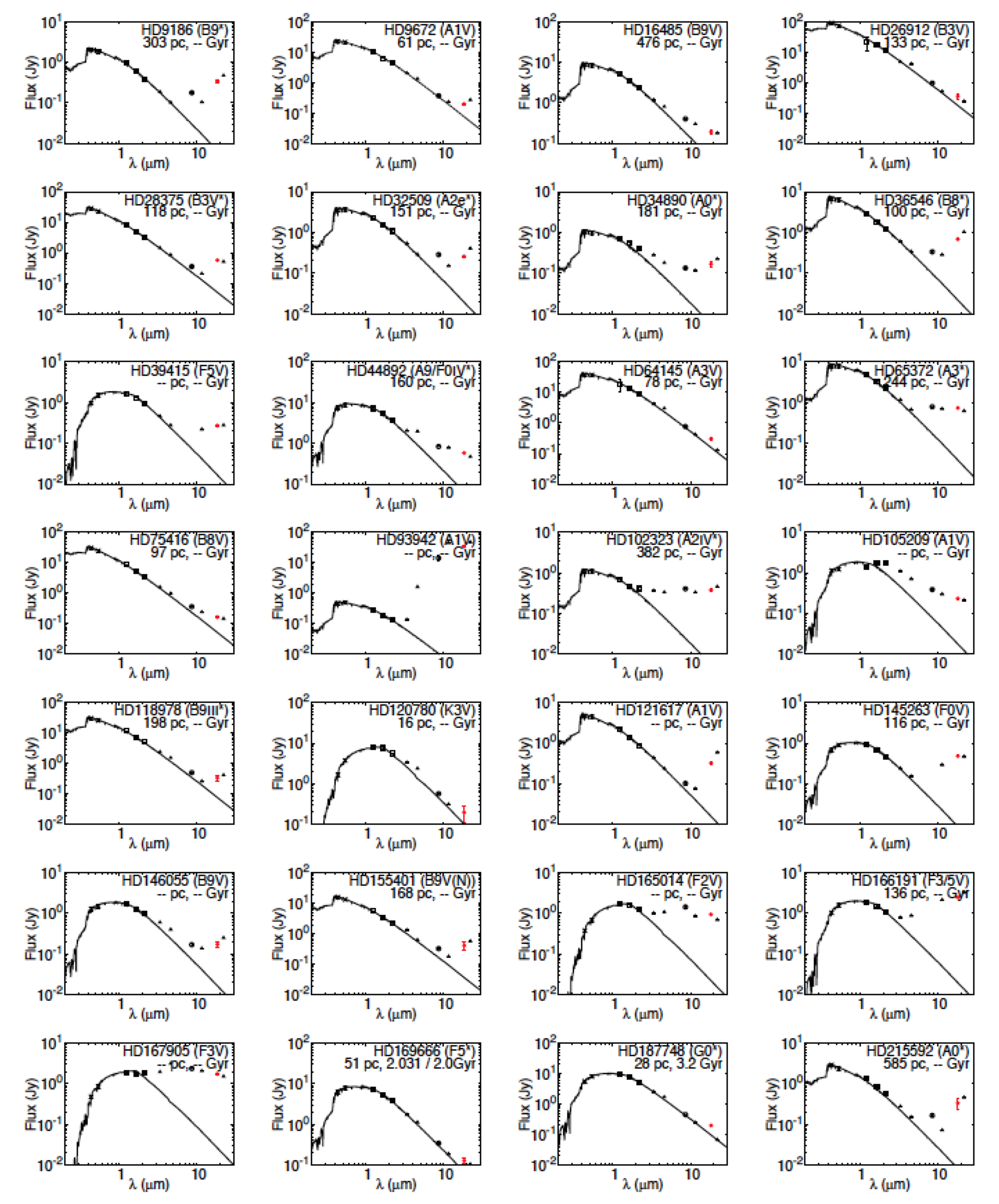}
\caption{Optical to mid-IR SEDs of our debris-disk candidates 
with 18\,$\mu$m excess emission.
Crosses, open squares, filled squares, open circles, filled circles, and triangles indicate
the photometric data points measured 
with Hipparcos (V$_{\rm T}$, B$_{\rm T}$) and USNO-B (I, z), 2MASS (J, H, Ks), 
IRSF (J, H, Ks), 
AKARI 9\,$\mu$m, AKARI 18\,$\mu$m, and WISE (3.4, 4.6, 12, 22\,$\mu$m), respectively. 
The red filled circles indicate AKARI 18\,$\mu$m flux used for excess identification.
Solid curves indicate the contribution of the photosphere
estimated on the basis of the optical to near-IR fluxes of the objects.
%
%
}\label{fig:SEDs}
\end{figure*}

\addtocounter{figure}{-1}
\begin{figure*}
\centering
\includegraphics[width=17cm]{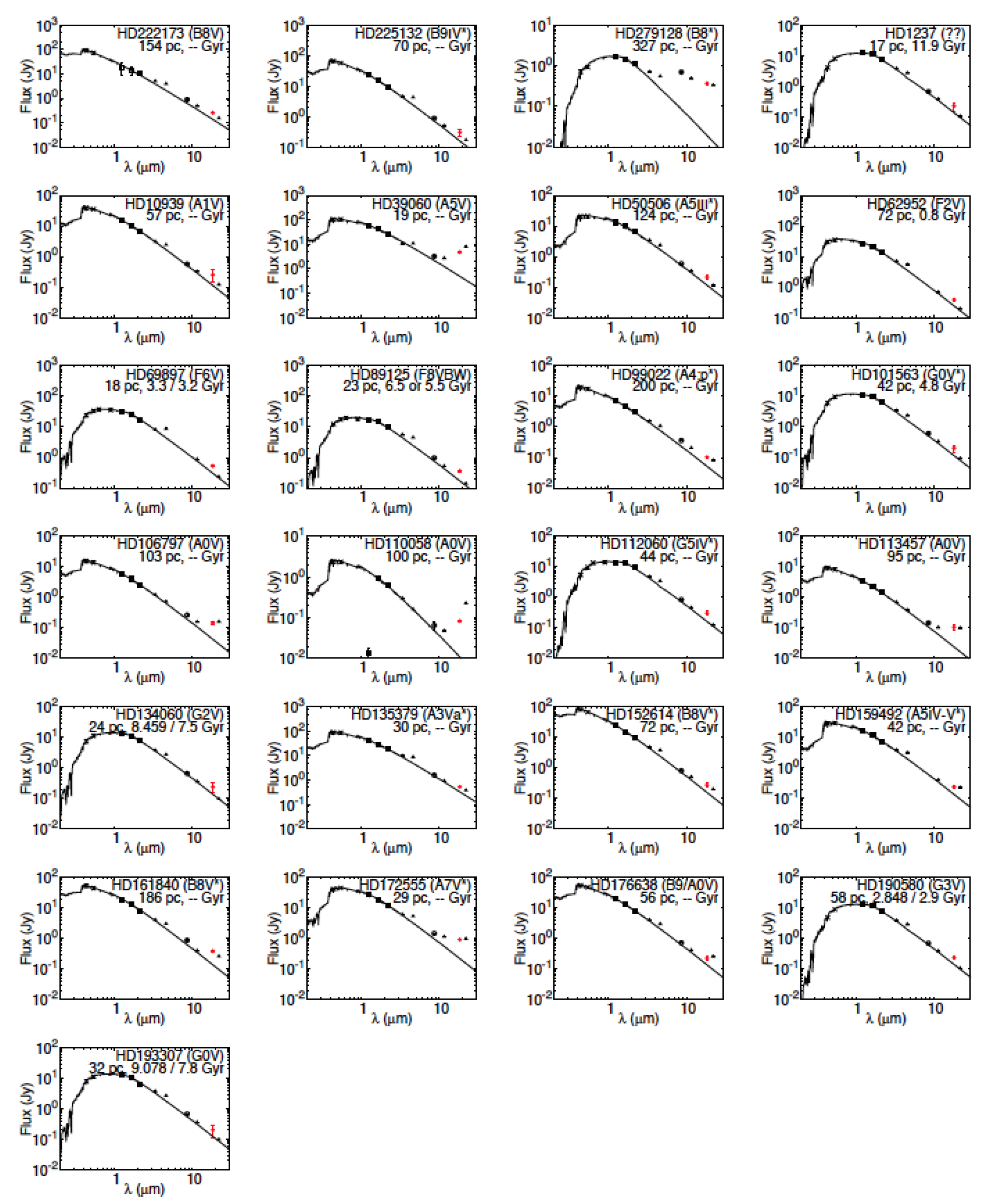}
\caption{Continued.}
\end{figure*}

\subsection{Reliability of mid-IR excess identification}
\label{reliability}

\begin{figure}[!h]
\centering
\includegraphics[width=6cm]{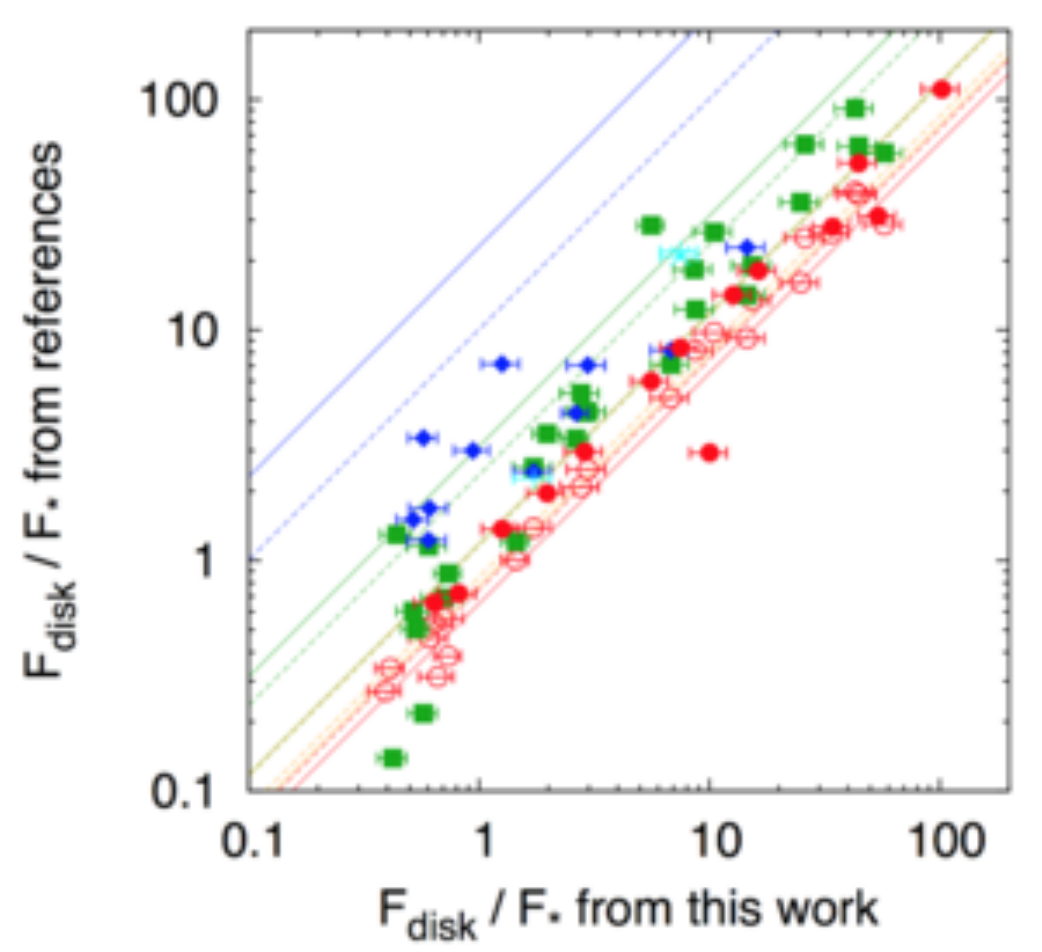}
\caption{
Flux ratio ($F_{\rm disk}/F_{*}$) in the previous works
plotted against the
flux ratio (for the AKARI 18\,$\mu$m) in this work for the same stars.
The filled circles,
open circles,
filled squares,
filled diamonds,
pluses,
crosses,
open triangles,
and open squares indicate
the flux ratios for the
AKARI 18\,$\mu$m data by \citet{Fujiwara13}, 
those by \citet{McDonald},
the WISE 22\,$\mu$m data by \citet{McDonald},
the IRAS 25\,$\mu$m data by \citet{McDonald},
the Spitzer/MIPS 24\,$\mu$m data by \citet{Su},
those by \citet{Rieke},
those by \citet{Trilling},
and those by \citet{Bryden06}, respectively.
For reference, the red, orange, yellow, green, and blue lines indicate
locations in this plot when the disk emission is 
the blackbody with $T$ = 800, 400, 200, 100, and 50\,K, respectively.
The dashed lines correspond to the Spitzer 24\,$\mu$m flux, and dash-dot-dashed lines
correspond to the WISE 22\,$\mu$m flux.
}
\label{fig:cmp}
\end{figure}

Among the \ndebris{} debris-disk candidates,
\prevdebris{} objects have been reported 
as debris disks in the previous studies 
\citep{Rieke,Bryden06,Su,Trilling,Fujiwara13},
and the other \conf{} objects have been reported 
as mid-IR excess candidates \citep{Oudmaijer,Clarke,McDonald}.
For evaluating the reliability of our excess estimate, 
we compare the excess ratio in our results 
with those in the previous works in Fig.~\ref{fig:cmp}
though the observed wavelengths are not exctly the same.
Figure~\ref{fig:cmp} indicates that
our estimate of the 18\,$\mu$m excess is 
consistent with the results in the previous works.
%
Excess ratios at the WISE\,22\,$\mu$m and IRAS\,25\,$\mu$m bands
tend to be larger than those at the AKARI\,18\,$\mu$m band,
which is reasonable if these systems have circum-stellar dust 
with temperatures lower than \chui{$\sim$300\,K}.
It also confirms that at least our excess ratios are not overestimated.

The available measurements with WISE and IRAS are also overlaid
on the individual SEDs in Fig.~\ref{fig:SEDs}.
Some stars, \chui{HD\,225132, HD\,1237, HD\,9186, HD\,10939},
show discrepancy 
between the AKARI and WISE-based measurements.
It could be attributed to
a temporal variation of the dust emission between 2006 and 2010.
Another possibility is an effect of the silicate emission features
that have broad peaks at 9 and 18\,$\mu$m.
We will address the nature of these objects in future work.

\subsection{Characteristics of our sample}
\label{char}

\begin{figure}[!h]
\includegraphics[width=8.5cm]{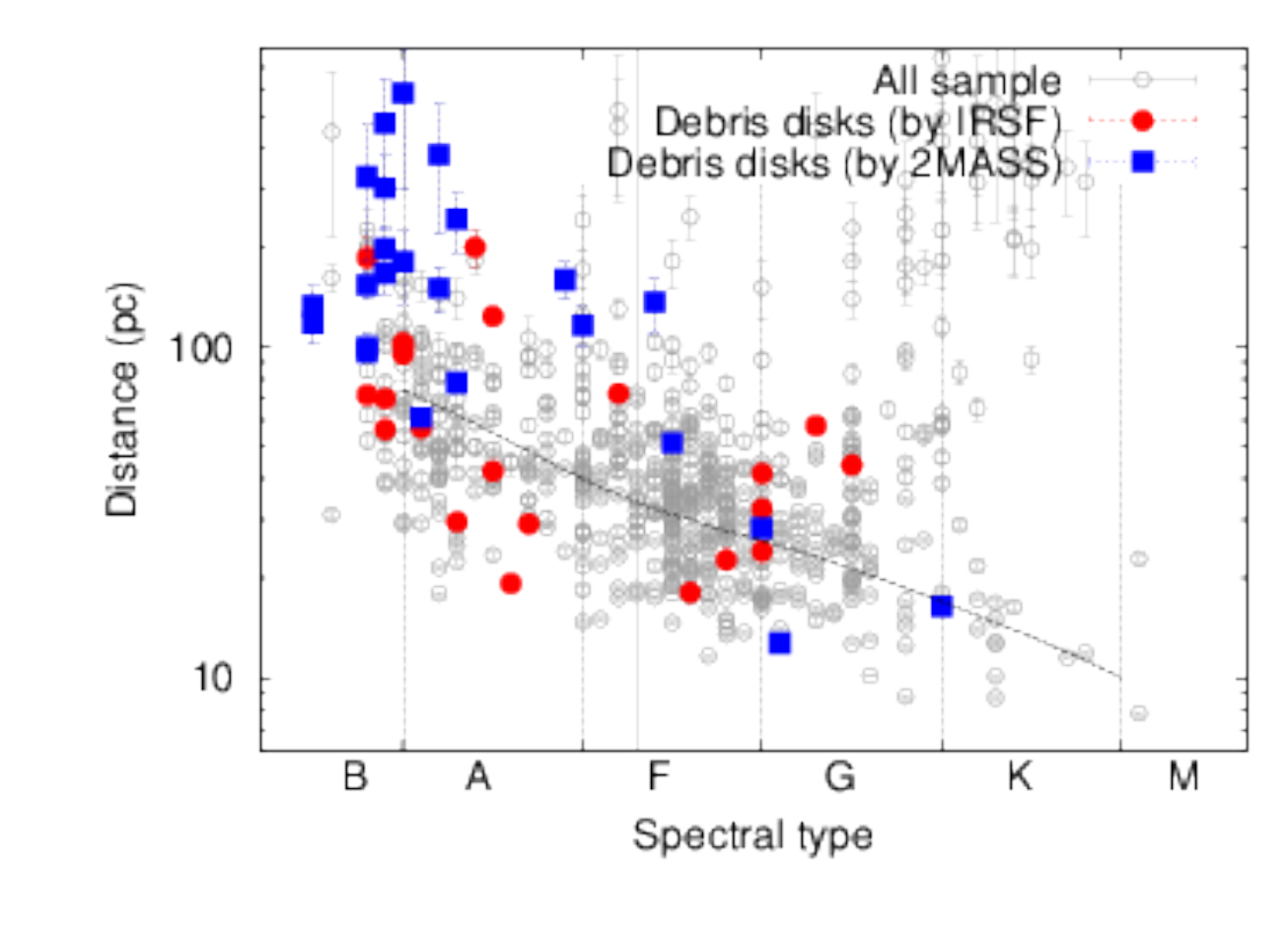}
\caption{Distance plotted as a function of the spectral type for our sample.
The open circles indicate all the 678 sample.
The filled circles and filled boxes indicate
the debris-disk candidates with the IRSF and 2MASS measurements, respectively.
The dotted curve indicates the detection limit for objects
without IR excess according to our definition.}
\label{fig:distance}
\end{figure}

Figure~\ref{fig:distance} shows the 
distance versus spectral type for our sample.
%
The detection limit and survey depth, is a function of the spectral type,
which is determined by the sensitivity of the AKARI mid-IR PSC.
\chui{From our sample and the detections shown 
in Fig.~\ref{fig:distance}, 
the survey depth for 3$\sigma$ detections of the photosphere reaches a distance of
\chui{74}\,pc for A0-type stars 
and \chui{10}\,pc for M0-type stars.}
The detection rate of debris disks for the AKARI--2MASS sample
is 7.9\,\% (31 objects out of 392).
That for the AKARI--IRSF sample, which covers nearby bright stars,
is 7.7\,\% (22 objects out of 286), which is comparable to that for the AKARI--2MASS sample.
If we use 2MASS fluxes for the AKARI--IRSF sample,
the detection rate of debris disks was 2.8\,\% (8 objects out of 286). 
The IRSF measurements significantly improve the detection rate.

%
%

Tables~\ref{tbl:list} and \ref{tbl:list2} list the debris-disk candidates detected by AKARI, 
which include previous disk detections. 
As shown in the list, 
\newobj{} objects are new detections and 
\conf{} objects are confirmation of 
the previous reports for IR excess detection
\citep{Oudmaijer,Clarke,McDonald}.
In our sample with 2MASS photometry,
newly detected objects 
around B-type stars (HD~146055) 
were not often explored in previous studies.
Our accurate determination of photospheric emission
by the IRSF observations results in 
new debris-disk detection around nearby bright F, and G-type stars
(HD~69897, HD~101563, HD~112060, HD~134060, and HD~193307). 
Therefore our sample contains mostly faint warm disks 
around bright nearby field stars.

\section{Discussion} \label{discussion}

\subsection{Debris-disk detection rate} \label{freq}

Table~\ref{tbl:freq} summarizes the total number of our main-sequence stars,
the number of debris-disk candidates, 
and the debris-disk detection rate for each spectral type.
\warn{The debris-disk frequency} varies smoothly 
from \chui{13}\,\% of the A-type to \chui{2}\,\% of the K-type sample 
though 
\warn{the numbers within each spectral-type sub-sample vary widely.
These debris-disk frequencies} 
are comparable to those reported in previous studies 
\citep{Rieke,Beichman05,Beichman06,Su,Thebault}.
The trend of increasing debris-disk frequency toward earlier types
is common among the unbiased volume limited surveys.

The number of main-sequence stars in our sample 
is largest for F-type stars
and decreases towards earlier-type stars (A- and B-type stars)
and towards later-type stars (G-, K-, and M-type stars).
The reason for this trend is explained as follows:
In a sensitivity limited unbiased survey,
early-type stars can be explored to a farther distance
than late-type stars
because early-type stars are brighter.
On the other hand, the number density of late-type stars is larger
than that of early-type stars,
according to the initial mass function of the solar neighborhood.
The number distribution of our sample
is the result of these two effects.

\begin{table*}[!t]
\caption{Statistics of debris-disk candidates 
with significant detection of the AKARI 18\,$\mu$m excess emission.}
\label{tbl:freq}
\begin{center}
\begin{tabular}{lrrrrrrr}\hline\hline
Spectral type&
\multicolumn{1}{c}{B8--9}&
\multicolumn{1}{c}{A0--9}&
\multicolumn{1}{c}{F0--9}&
\multicolumn{1}{c}{G0--9}&
\multicolumn{1}{c}{K0--9}&
\multicolumn{1}{c}{M0--9}&
\multicolumn{1}{c}{Total}\\
\hline
Number of stars                 & 35 & 150 & 280 & 156 &  54 & 3 & 678\\
Number of debris-disk candidates& 15 &  21 &   9 &   7 &   1 & 0 &  53\\
Detection rate ($\%$)           & 43 &  14 &   3 &   4 &   2 & 0 &   8\\
\hline
\end{tabular}\\
\end{center}
\end{table*}

\subsection{Disk dissipation timescale} \label{dissipation}

Small grains, which contribute most to infrared emission,  
are removed by collisional fragmentation
and blown out by radiation pressure. The removal timescale is much
shorter than the ages of host stars. 
Disruptive collisions among underlying large bodies, 
which are called planetesimals, produce smaller bodies and collisional
fragmentation among them results in even smaller bodies. 
This collisional cascade continues to supply small grains. 
The evolution of debris disks has been explained 
by the steady-state collisional cascade model 
\citep[e.g.,][]{Wyatt08,Kobayashi10}: the total mass of bodies decreases 
inversely proportional to time $t$. 
Therefore, the excess ratio ($F_{\rm disk}/F_{*}$) is 
given by 
\begin{equation}
\frac{F_{\rm disk}}{F_{*}} = \frac{t_0}{t},\label{cc} 
\end{equation}
where $t_0$ is the dissipation timescale that is determined by
the collisional cascade. 
Under the assumption of the steady state of collisional cascade, the
power-law size distribution of bodies is analytically obtained and the
power-law index depends on the size dependence of the collisional
strength of bodies \citep[see Eq.~(32) of ][]{Kobayashi10}. In the obtained size
distribution, erosive collisions are more important than catastrophic
collisions \citep[see Fig.~10 of][]{Kobayashi10}. Taking into account the size
distribution and erosive collisions, we derive $t_0$ according to the
collisional cascade (see Appendix \ref{appendix_col} for derivation),
\begin{eqnarray}
t_0\sim 1.3 \left( \frac{s_{\rm p}}{\rm 3000\,km} \right)^{0.96}
\left( \frac{R}{\rm 2.5\,AU} \right)^{4.18}\nonumber\\
\times \left(\frac{\Delta R}{0.4 R}\right)
\left( \frac{e}{\rm 0.1} \right)^{-1.4} {\rm Gyr},\label{eq:t0}
\end{eqnarray}
where $s_{\rm p}$ is the size of planetesimals, 
$R$ is the radius of the planetesimal belt, 
and $e$ is the eccentricity of planetesimals. 
Interestingly, $t_0$ is independent of the initial number density of planetesimals \citep{Wyatt07}.
Note that 
the perturbation from Moon-sized or larger bodies is needed 
to induce the collisional fragmentation of planetesimals
\citep{Kobayashi}, which is implicitly assumed in this model. 

\begin{figure*}[H]
\includegraphics[width=8.5cm]{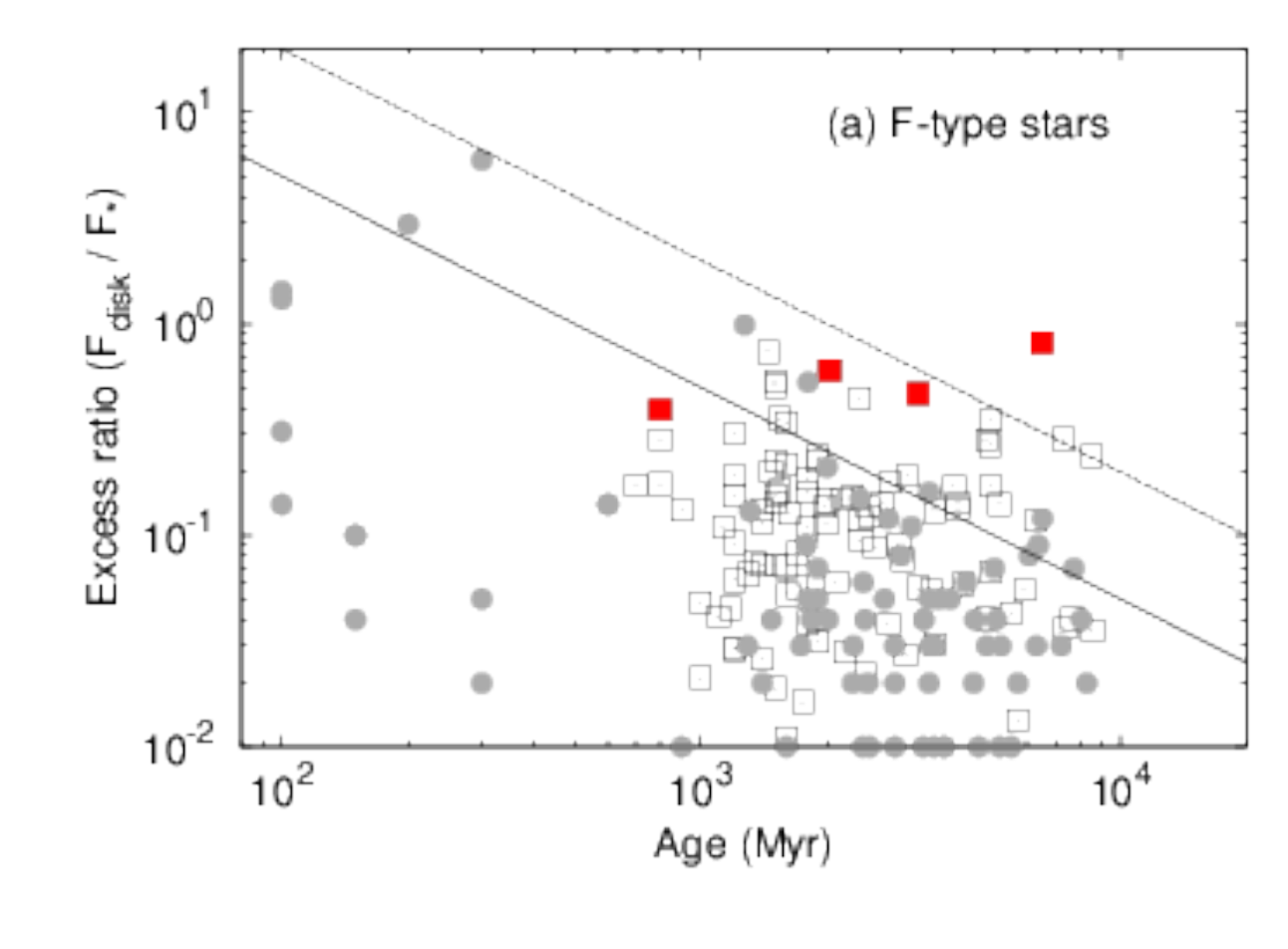}
\includegraphics[width=8.5cm]{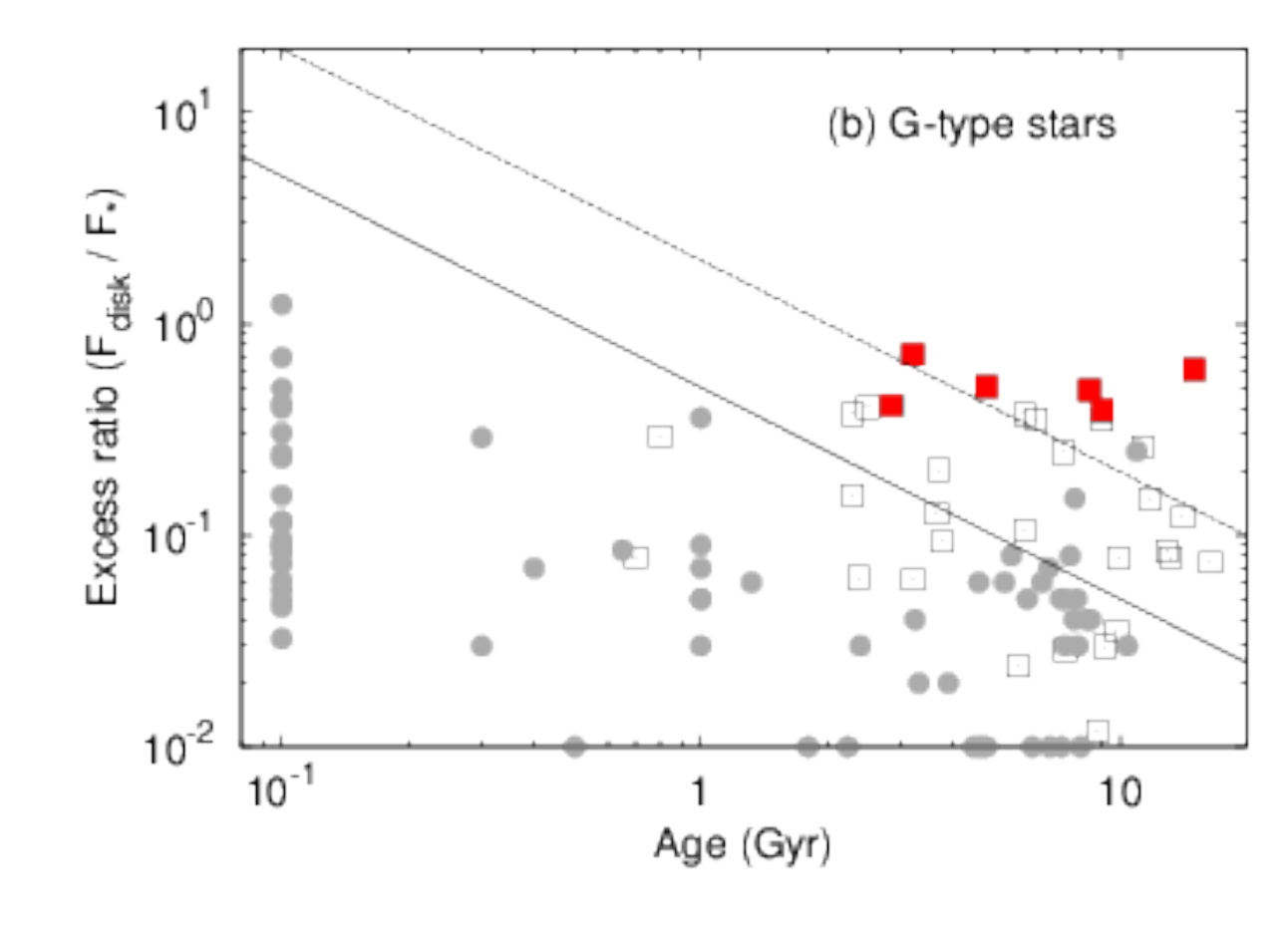}
\caption{
(a) $F_{\rm disk}/F_{\ast}$ at AKARI 18\,$\mu$m versus stellar age for F-type stars.
Filled circles indicate debris disk samples from previous works
at 24\,$\mu$m by Spitzer/MIPS
\citep{Beichman05,Beichman06,Bryden06,Chen05A,Chen05B,Hillenbrand,Trilling}.
Filled squares represent excess ratios observed by AKARI at
18\,$\mu$m for stars with excesses larger than $3\sigma$ (debris
disk candidates), while open squares show that for all stars in our
sample.  
The solid line indicates the evolutionary track with $t_0=0.5$\,Gyr,
where $t_0$ is the dissipation time scale,
while the dotted line indicates evolutionary track of $t_0=2$\,Gyr 
(see text for details).
(b) Same as (a) but for G-type stars.
}
\label{fig:dissipation}
\end{figure*}

Figure \ref{fig:dissipation} shows 
excess ratio, $F_{\rm disk}/F_{*}$, versus stellar age. 
We plot our samples if stellar ages are known: 
Estimated ages are available for four and six objects 
among nine F-type and seven G-type stars in Table 3 and 4, respectively (Chen et al. 2001; Feltzing et al. 2001; Holmberg et al. 2009).
We also plot the samples obtained 
from previous observations.  
The excess ratios for
most of the objects are explained by the steady-state collisional
cascade model (Eq.~\ref{cc}) if $t_0<0.5$\,Gyr.
%
However at least \nlarge{} objects in our sample, 
the ages of which are determined in the literature, 
can not be explained with $t_0 < 0.5$\,Gyr
and are rather consistent with $t_0$ of 2\,Gyr. 
If we assume a system like the solar system that has
$R \approx 2.5$\,AU and $e \approx 0.1$, then, 
very large planetesimals with $s_{\rm p} \sim 5,000$\,km are required
for $t_0\sim 2$\,Gyr (see Eq.~(\ref{eq:t0})). 
The bodies with $s_{\rm p}\sim 5,000$\,km are larger than the Mars. 
A small number of such large bodies can be formed but a swarm of such large bodies for
collisional cascade may be unrealistic.  Furthermore, there are
no young objects with high fractional luminosities corresponding
to $t_0$ longer than 2\,Gyr (see Fig.~\ref{fig:dissipation}), which are progenitors of
those old, bright debris disks. Therefore, those old debris disk
objects with high fractional luminosity may not be explained only
by the conventional steady-state cascade model. 

\citet{Kennedy} explored IR excess for Kepler Objects 
by using the WISE catalog and indicated that 
large excesses around in old stars can be explained 
by chance alignment of interstellar dust or background galaxies.
\citet{Merin} observed stars showing warm-IR excesses in 
WISE bands 3 and 4 with Herschel and obtained no detection 
in any of the targets, which indicates most of such excesses 
are likely caused by chance alignment of the foreground or 
background objects. Though our sample covers brighter nearby 
objects than the distant WISE sample, we investigate 
the probability of chance alignment of known extragalactic 
sources and/or diffuse dust emission by using NED database 
and the AKARI 18\,$\mu$m images, respectively. 
No suspected features are found in these processes (see Section \ref{sec:excess}). 
Therefore, we judge that the stars have debris disks. 
High spatial resolution observations with future large telescopes 
might resolve the disk or reveal a nearby background source, 
thus clarifying the origin of the excesses.

If they are true debris disks, these
high excesses around old stars may be related to other
non-steady processes such as follows. (a) In planet formation, a
swarm of planetesimals produces a small number of planetary
embryos and the stirring by planetary embryos induces collisional
fragmentation of remnant planetesimals, resulting in the
formation of debris disks: Low mass disks composed of large
planetesimals tend to have long timescales of disk
evolution \citep{Kobayashi}, which may explain these high
excess around old stars. (b) In the late stages of planet
formation, giant impacts among Mars-sized planetary embryos, which
produced the Moon in the solar system, occur through long-term
orbital instability. Although the total mass of fragments ejected
from giant impacts is much smaller than planetary embryos, the
excesses ratios resulting from giant impacts increase to
observable levels \citep{Genda}, which may form late
debris disks. (c) In the solar system, the late heavy bombardment
is believed to have occurred at $\sim3.9$\,Gyr, based on radiometric
ages of impact melts of lunar samples \citep{Tera}.  It may be related to
dynamical events of planets in the solar system, which induce the
formation of late debris disks in other systems \citep{Booth,Fujiwara13}.
(d) In planet-hosting systems,
planets trap dust grains in their mean motion resonances in a
long timescale 
\citep{Liou96}, which may form late bright
debris disks. The resonance trap is particularly studied for the
Earth’s resonance orbits in the solar system by the past infrared
survey missions \citep[e.g.,][]{Kelsall,BMay}.
Furthermore, the temporal variation of this component is
indicated by the recent analysis of the AKARI all-sky survey data
\citep{Kondo}, although the variability is small.

Each non-steady process leads to different temporal evolution of excess
ratios. The origin of the high excess debris disks around old stars
will be revealed by investigating the temporal variability of infrared
excess emission via multi-epoch observations.  The next chance will be
brought us by the next space infrared mission, JWST or SPICA.  
Imaging observations with ground-based large telescopes such as TMT are also expected.
%
%
The first detections of such planets are being made by 
ground-based direct imaging surveys, 
and space-based detections will follow in the future (e.g., WFIRST).

\section{Summary}
By using the AKARI mid-IR all-sky PSC,
we have explored debris disks with 18\,$\mu$m excess emission.
We have carefully selected nearby isolated stars
and compared their estimated photospheric fluxes
with observed fluxes at a wavelength of 18\,$\mu$m.
For accurate estimation of the photospheric fluxes of the central stars,
we have performed J, H, and Ks band photometry with IRSF
for nearby bright stars
whose 2MASS fluxes have large uncertainties due to saturation.
The flux uncertainties of the central stars have been improved
from \aveimprove{} on average.
As a result, we have successfully detected \ndebris{} debris-disk candidates 
out of \total{} main-sequence stars.
At least \nlarge{} objects of them 
have large excess emission for their ages, 
which cannot be explained 
by the conventional steady state collisional cascade model.

\begin{acknowledgements}
This research is based on observations with AKARI, 
a JAXA project with the participation of ESA.
The IRSF project is a collaboration between Nagoya University and the
South African Astronomical Observatory (SAAO) supported by the
Grants-in-Aid for Scientific Research on Priority Areas (A) (No.
10147207 and No. 10147214) and Optical \& Near-IR Astronomy
Inter-University Cooperation Program, from the Ministry of
Education, Culture, Sports, Science and Technology (MEXT) of Japan and
the National Research Foundation (NRF) of South Africa.
This publication makes use of data products 
from the Two Micron All Sky Survey, 
which is a joint project of the University of Massachusetts 
and the Infrared Processing and Analysis Center/California 
Institute of Technology, funded by the National Aeronautics 
and Space Administration and the National Science Foundation.
This research has made use of the NASA/IPAC Infrared Science Archive,
and the NASA/IPAC Extragalactic Database (NED),
which is operated by the Jet Propulsion Laboratory, 
California Institute of Technology, 
under contract with the National Aeronautics and Space Administration. 
This research has made use of the SIMBAD database, 
operated at CDS, Strasbourg, France.
We thank Dr. G. Kennedy for careful reading and emendation of the manuscript.
This work is supported by 
the Grant-in-Aid for the Scientific Research Funds
(No.~24740122, No.~26707008, No.~50377925, and No.~26800110) from 
Japan Society for the Promotion of Science (JSPS),
and (No. 23103002) from the Ministry of Education, Culture, Sports, Science and Technology (MEXT) of Japan.
\end{acknowledgements}

\newpage
\begin{appendix}
\section{Accurate J, H, Ks photometry of nearby bright stars}\label{A1}
Using the IRSF telescope \citep{Sato},
we have performed J, H, and Ks band photometry of \nIRSF{} bright stars
which have large photometric uncertainties due to saturation 
in the 2MASS catalog \citep{Cutri}.
IRSF is the near-infrared telescope with
a $\phi$1.4\,m primary mirror located in Sutherland, 
South Africa at 1,800\,m elevation. 
It is managed by Nagoya University. 
SIRIUS \citep{Nagayama} is the wide-field camera, 
which enables simultaneous J, H, and Ks band wide field imaging. 
The field-of-view (FOV) size is $7.'8\times7.'8$. 
The pixel scale is $0.''45$, while
the PSF size is $1''$--2$''$ depending on the weather. 

\begin{figure}[!h]
\center
\includegraphics[width=8cm]{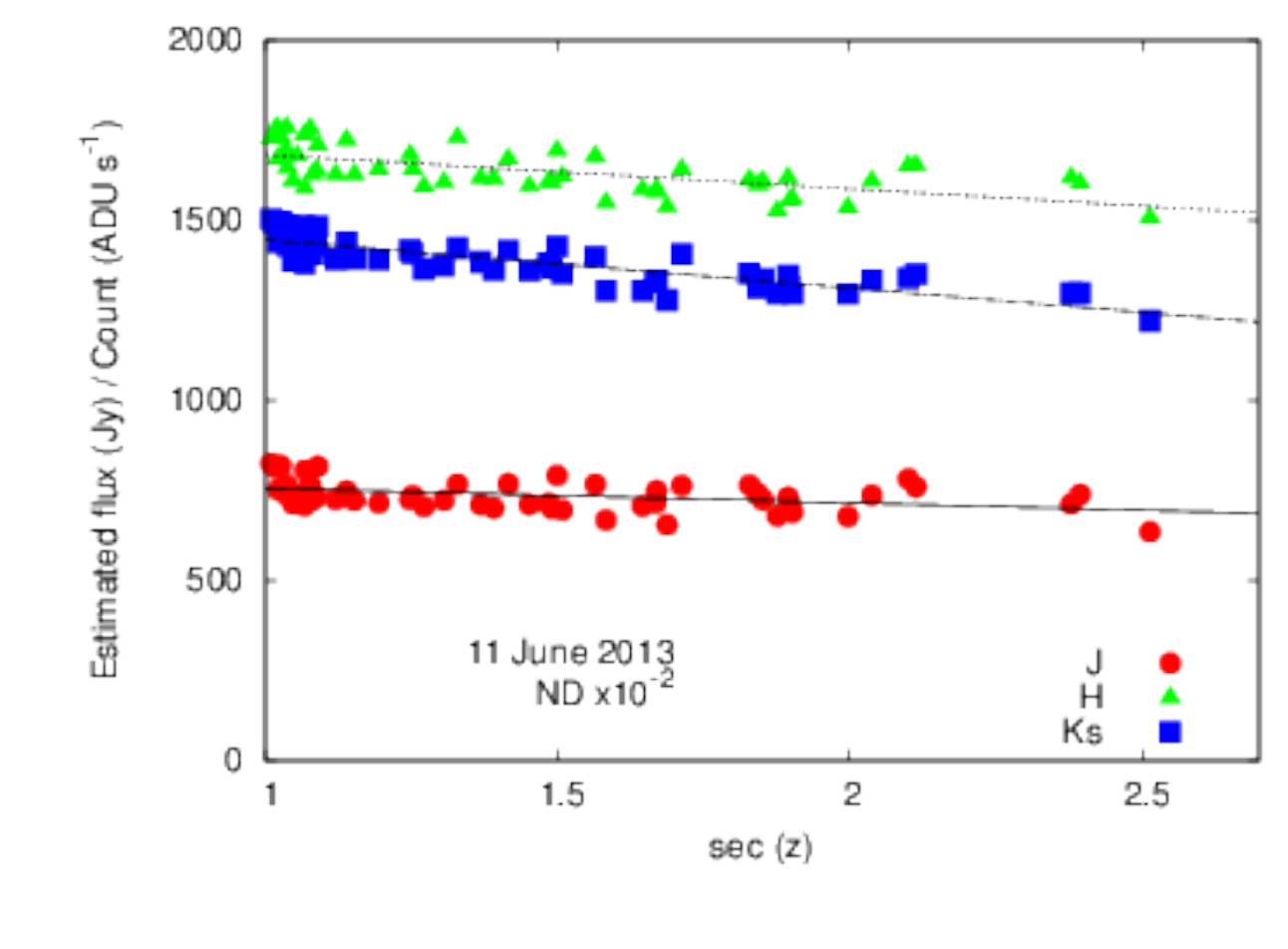}
\caption{Ratio of the estimated fluxes over the observed fluxes of standard stars
as a function of $\sec(z)$. This example is of the observation on 11th June 2013.
This plot is made for every night for every ND filters.
The circle, triangle, and square indicate the J, H, and Ks bands, respectively.
The solid line, the dotted line, and the dot-dashed line indicate
fitting results for the J, H, and Ks bands, respectively.
}
\label{fig:secz}
\end{figure}

The observations were carried out 
using the ND filter for $\times10^{-2}$ flux attenuation
in six nights with relatively stable weather: 
2011 August 6, 2011 August 9, 2012 February 5, 2012 February 10,
and 2013 June 12.
%
In the observations using ND filters, 
we cannot make flux calibration using standard stars in 
the same FOV,
because, in most cases, only the target star is detected in each image. 
Based on the observations of bright standard stars \citep{Carter},
we derive the system response (estimated flux / observed flux (count s$^{-1}$)) 
as a function of $\sec(z)$
specific to each night and each ND filter
where $z$ is a zenith angle.
A set of examples of the functions for the J-, H-, and Ks-bands 
in a night are shown in Fig.~\ref{fig:secz}.
The system response is determined with an accuracy of 0.1--0.2~\%.
Then we applied these functions to each observation.

Observational data are reduced 
with the standard pipeline for the SIRIUS data.
The images are stacked and aperture photometry is 
employed for each target star 
using the IRAF phot package \citep{Tody}.
The parameters are optimized 
to maximize the signal-to-noise ratios and 
to obtain total fluxes of stars without an aperture correction. 
We adopted an aperture radius of 5$''$, an annulus
(the width of the gap between the source area and the sky area) of 3$''$ ,
and a sky width (the width of the annulus of the sky area) of 3$''$.

\begin{figure*}[!t]
\includegraphics[width=18.5cm]{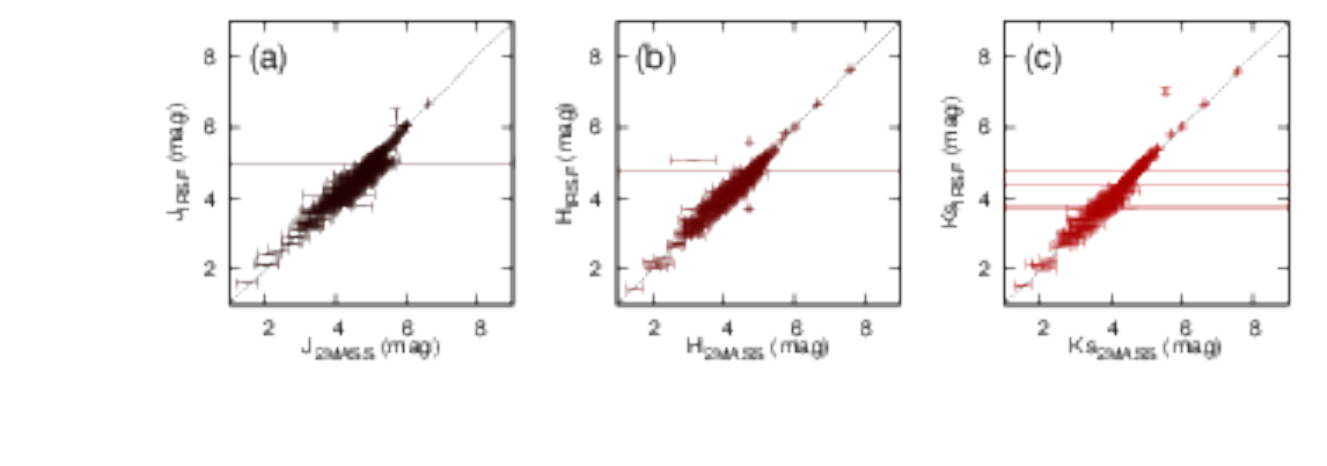}
\caption{IRSF photometry versus 2MASS photometry for our sample of \nIRSF{} 
bright main-sequence stars in the J band (a), H band (b), and Ks band (c).}
\label{fig:2MASSvsIRSF}
\end{figure*}

\begin{figure*}[!t]
\center
\includegraphics[width=18.0cm]{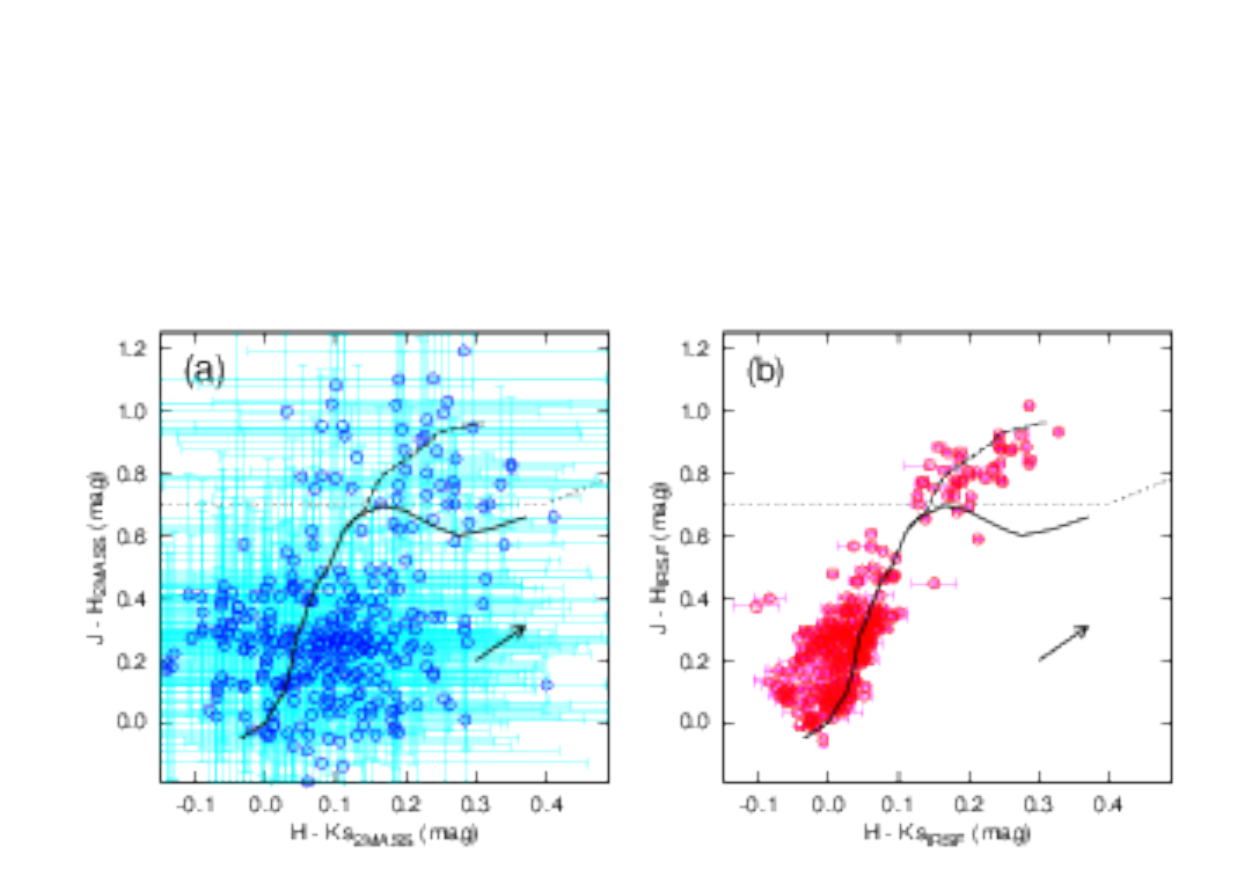}
\caption{
(a) The J$-$H versus H$-$Ks color-color diagram 
of bright main-sequence stars
based on the 2MASS measurements.
The solid curve indicates locus of main-sequence stars 
while the dotted curve indicates that of giant stars \citep{Bessell}.
The solid arrow shows the interstellar extinction vector 
for Av$=1$ mag, using the \citet{Weingartner} Milky Way model of Rv$=3.1$.
(b) Same as (a), but for the IRSF measurements.
%
\chui{The objects above dashed lines are removed from our main-sequence 
sample because they might be giant stars.}
}
\label{fig:JHK}
\end{figure*}

Table~\ref{tbl:jhk} summarizes J, H, Ks photometric results 
for our target stars. 
Figure~\ref{fig:2MASSvsIRSF} compares the IRSF J, H, Ks fluxes 
with the 2MASS J, H, Ks fluxes for the same stars. 
Our IRSF measurements are statistically consistent 
with the 2MASS measurements within the uncertainties.
The error bars along the horizontal axis are systematically longer than 
those along the vertical axis,
indicating that 
the IRSF measurements have smaller uncertainties 
than the 2MASS measurements. 
The relations between the flux and flux errors are shown in Fig.~\ref{fig:err}.
The averaged flux errors are reduced 
from 17\%, 14\%, and 11\% to 1.9\%, 1.4\%, and 2.0\%, for the J, H, and Ks band, 
respectively.
%
In Fig.~\ref{fig:JHK}, 
we compare the color-color diagram based on the 2MASS photometry 
and that based on the IRSF photometry.
While the 2MASS measurements show a large scatter,
our IRSF measurements trace the intrinsic locus of main-sequence stars
\citep{Bessell}.
This confirms the reliability of the IRSF measurements.


\section{Reliability of the fitting of photospheric emission}\label{A2}
\begin{figure*}[!t]
\includegraphics[width=8.5cm]{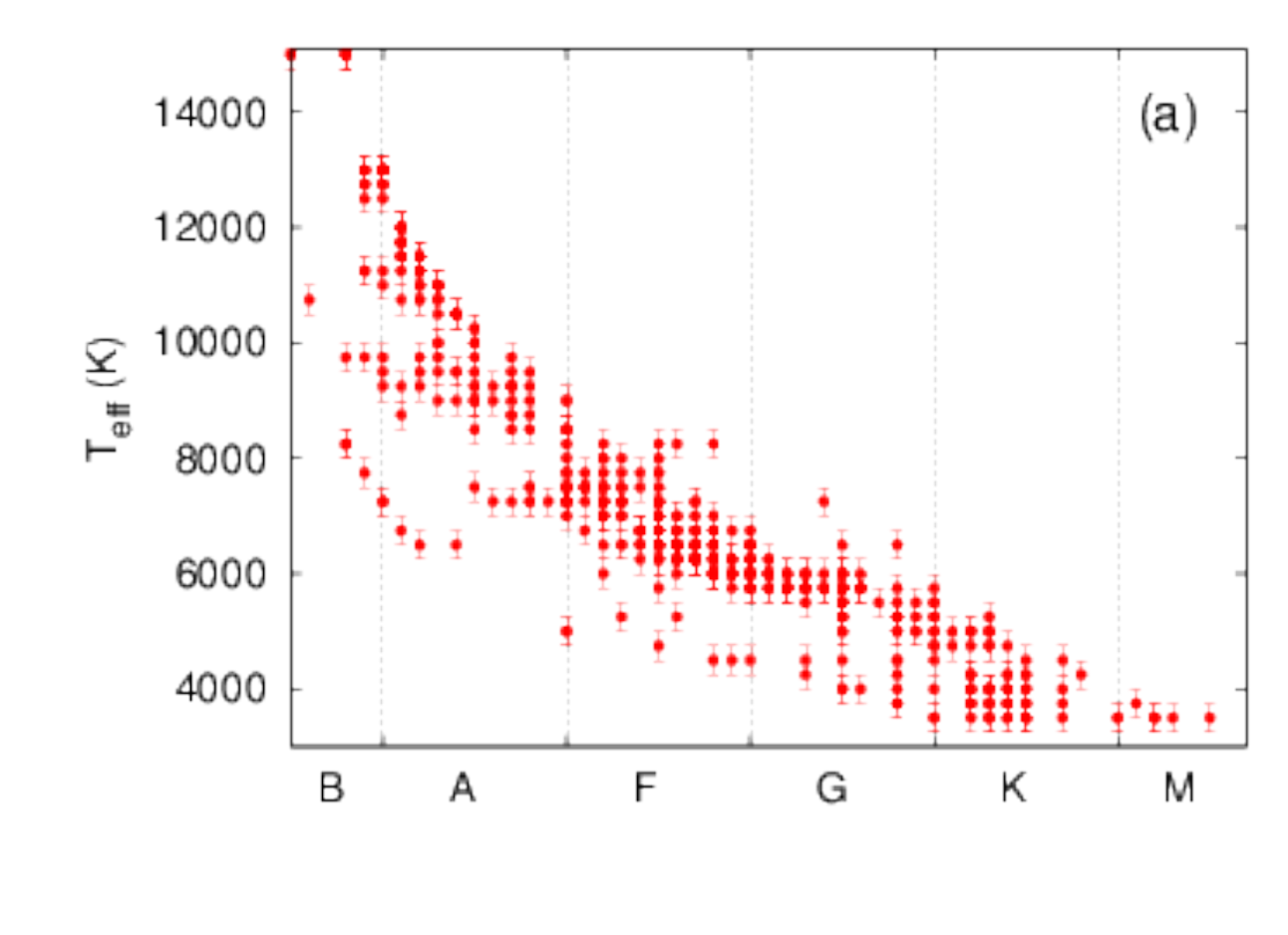}
\includegraphics[width=9cm]{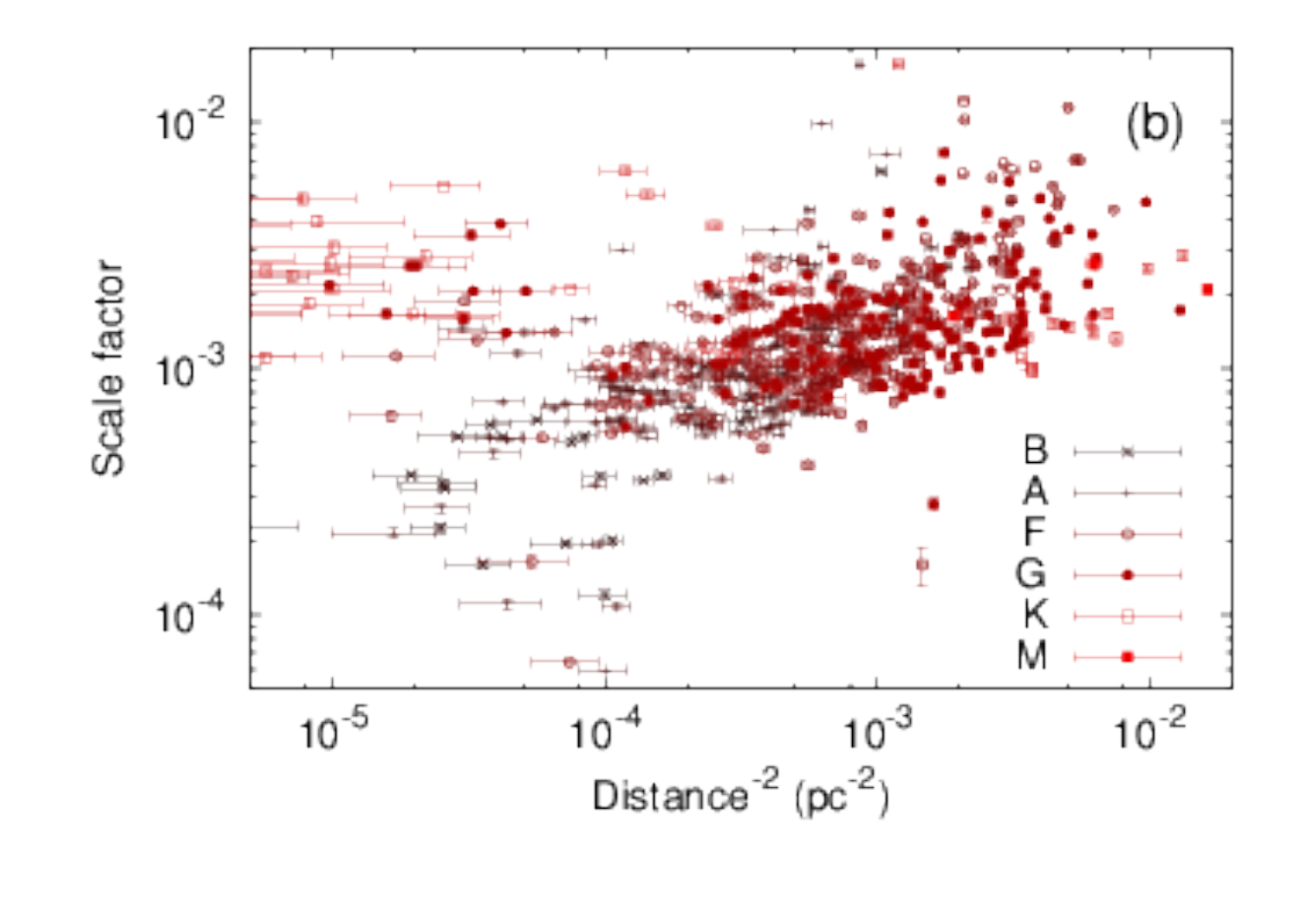}
\caption{(a) Effective temperature of a central star (fitting result)
versus spectral type from the literature \citep{Wright}.
(b) Scale factor $S$ (fitting result) versus
reciprocal of the square of the distance from
the Hipparcos catalog \citep{Perryman}.
}\label{fig:teff}
\end{figure*}

For the evaluation of the fitting results
in the estimation of photospheric emission,
we compare the output parameters ($T_{\rm eff}$ and $S$)
with the related parameters quoted from 
the literature (spectral type and distance).
Figure~\ref{fig:teff}a shows $T_{\rm eff}$ versus spectral type
and Figure~\ref{fig:teff}b shows $S$ versus distance$^{-2}$
for our sample.
All the objects are aligned along the intrinsic locus
of main-sequence stars in both plots.
These indicate that 
our sample selection and fitting process work well in general.

\section{Uncertainties in SED fitting of photosphere using pre-computed grids}\label{A2b}

In estimating 18\,$\mu$m photospheric emission of stars
by fitting their optical-to-near-infrared SEDs with the Kurucz model \citep{Kurucz},
we use the pre-computed grids for $T_{\rm eff}$, $\log(g)$, and metallicity.
By using these quantized parameters to the fitting,
the uncertainty for $T_{\rm eff}$ is certainly increased.
But its effect is small for the predicted $F_{*}$ at 18\,$\mu$m.
It is much smaller than the total systematic error 
considered in the excess identification:
0.126 for AKARI--IRSF sample and 0.182 for AKARI--2MASS sample (see Section~\ref{sec:excess}).
This is because the change in $T_{\rm eff}$ affects spectra in shorter wavelengths ($<1$\,$\mu$m)
and less affects longer wavelengths.

\begin{figure}
\includegraphics[width=9cm]{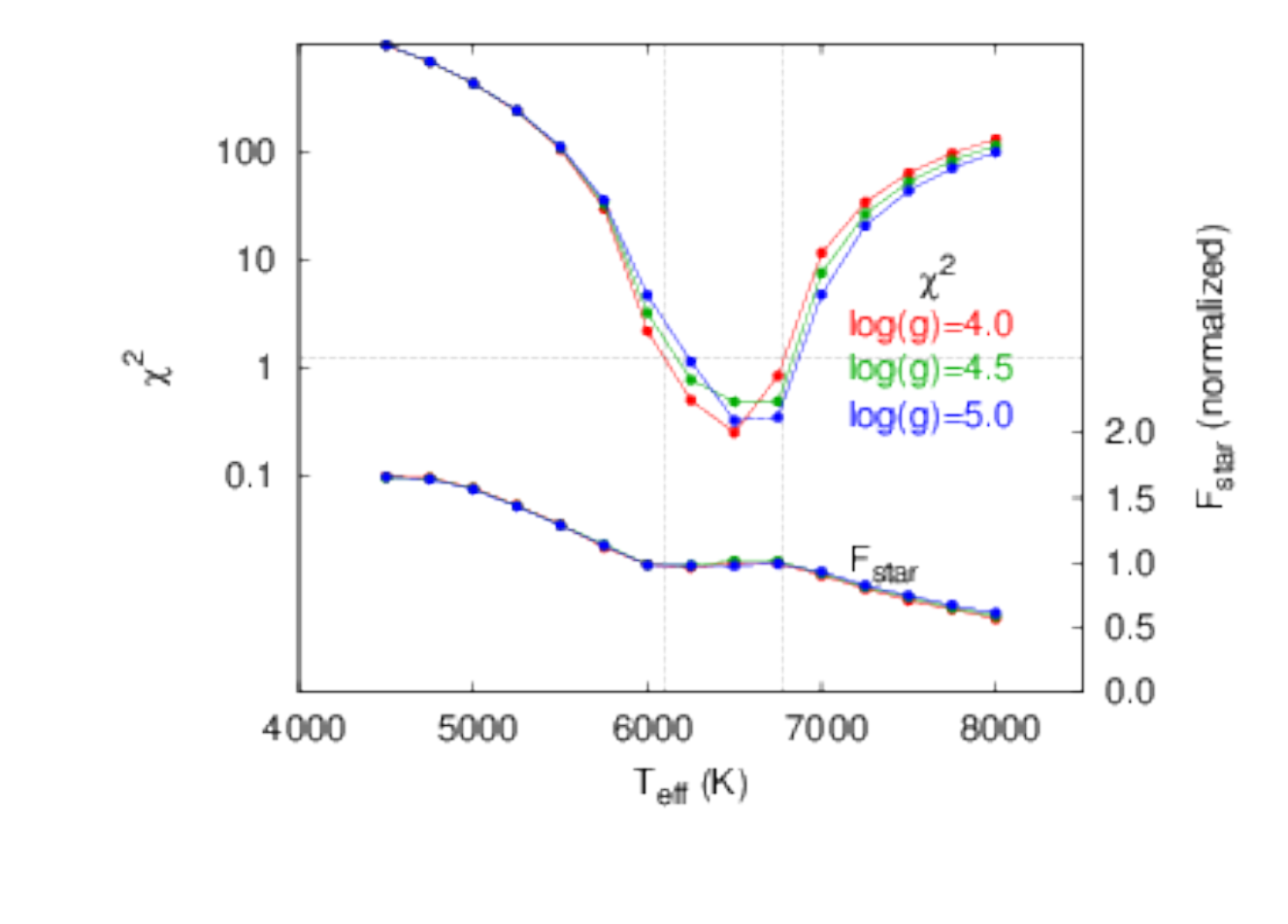}
\caption{$\chi^2$ versus $T_{\rm eff}$ in the fitting photospheric emission for HD\,187748 as an example (left axis).
Red, green, and blue points indicate
cases for $\log(g)=4.0$, 4.5, and 5.0, respectively.
Deviation of $F_{\rm *,18}$ around the best-fit value 
($F_{\rm *,18}$ at $T_{\rm eff}$=6,500, $\log(g)=4.0$), 
is also overlaid in this plot (right axis).
}\label{fig:chi2Teff}
\end{figure}

\begin{figure}
\includegraphics[width=9cm]{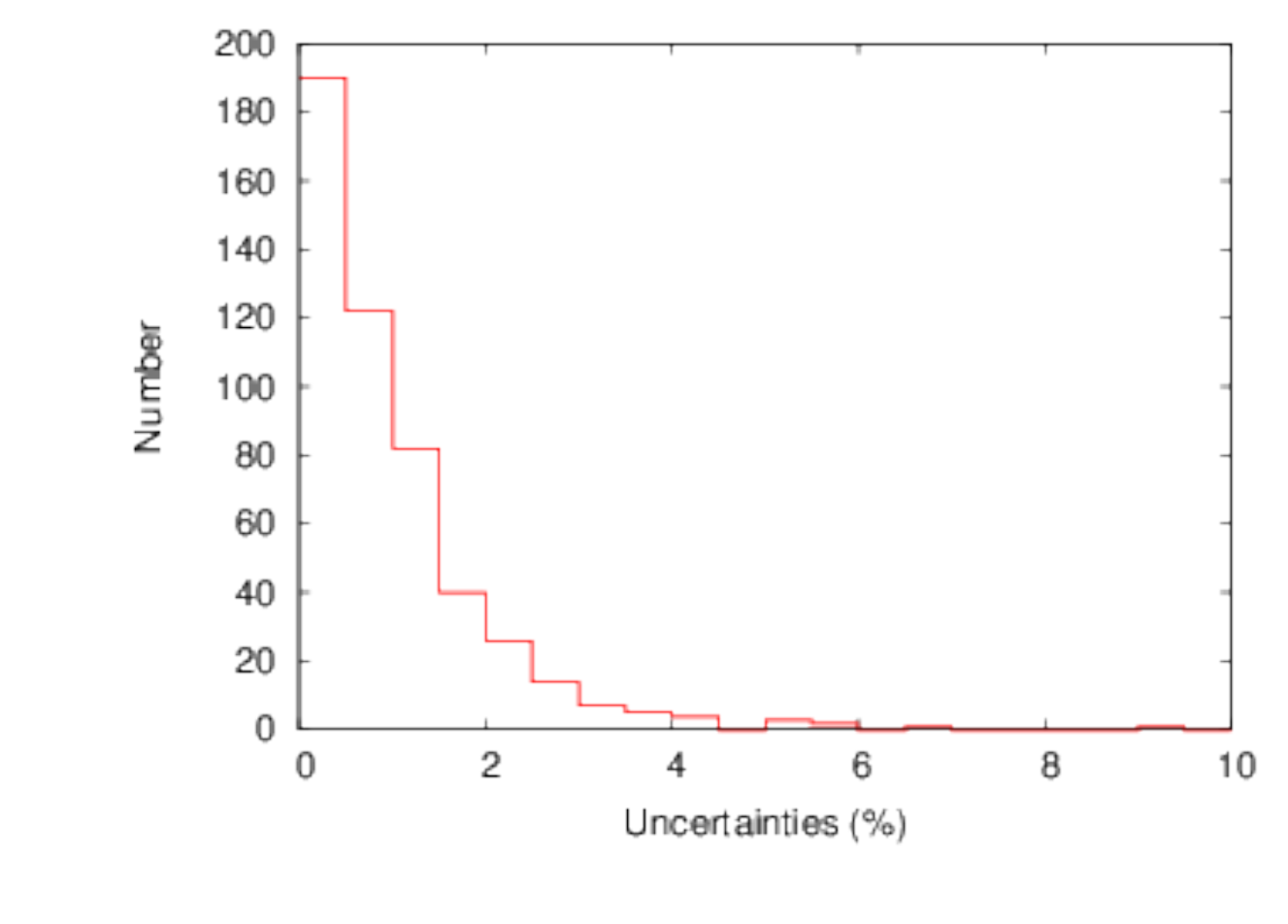}
\caption{Distribution of uncertainties in the photospheric fitting process.}
\label{fig:chi2stat}
\end{figure}

Fig.~\ref{fig:chi2Teff} shows an example of $\chi^2$ versus $T_{\rm eff}$ plot 
for the SED fitting of the star (HD\,187748).
The $T_{\rm eff}$ range which satisfies $\Delta\chi^2 = \chi^2 - \chi^2_{\rm min} < 1.0$
is 6140--6760\,K, which is the 68.3\% confidence range assuming the normal distribution, 
and thus 1$\sigma$ equivalent uncertainties.
If the sampling against $T_{\rm eff}$ is finer,
$\chi^2$ at the true $T_{\rm eff}$ might be smaller,
and the $T_{\rm eff}$ for 1$\sigma$ is smaller.
However, the uncertainty in the estimate of $F_{\rm star}$ is small enough
even in this $T_{\rm eff}$ range (0.4\% for this sample).
We estimate uncertainties in $T_{\rm eff}$ and $F_{*}$ for each star in this method.
Fig.~\ref{fig:chi2stat} shows the distribution of the 1$\sigma$ uncertainties
in the photospheric fitting for all the stars in our sample.
The uncertainties in the fitting process are much smaller than the
total systematic error considered in the excess identification.
Thus deviation of $T_{\rm eff}$ does not significantly affect the prediction of $F_*$.

It should be noted that
the fitting result of $T_{\rm eff}$ for earlier type stars tends to be
different from the value expected from the spectral type.
The $T_{\rm eff}$ for 
HD~75416 (B8), HD~161840 (B8), and HD~222173 (B8) result in 15,000\,K, the maximum value in the fitting range.
They do not 
show a convincing $\chi^2$ curve and an asymptotic trend to the value of 15,000\,K.
It might because the peak of the photospheric SED is shorter than the wavelengths of input data.
The $T_{\rm eff}$ for HD~279128 (B8), HD~146055 (B9) result in 
6,000\,K and 7,000\,K, respectively. 
It might be due to the constraint in $A_V$ in addition to the large 2MASS photometric errors.
However, they don't affect the excess identification
because all of them show apparent mid-IR excess in Fig.~3.

\section{Point image analysis for debris-disk candidates} \label{imageana}
%
%

We have investigated the images in 2MASS Ks-band, and AKARI 9\,$\mu$m,
as well as those in AKARI 18\,$\mu$m for all the 53
debris-disk candidates to check image artifacts and contamination
of other sources. Fig.~\ref{fig:cross-id} shows these images. 
In previous work looking for WISE infrared excesses around faint stars 
contamination and artifacts have posed problems \citep{Kennedy,Ribas}.
%
However, the fluxes of our debris-disk
candidates are at the brightest end of WISE dynamic range.

AKARI has better spatial resolution than WISE.
Thus, the AKARI images less suffer confusion than the WISE data.
In the AKARI 18\,$\mu$m images, effects of image artifacts 
were not found and contaminating sources were not recognized. 
For example, HD\,34890, HD\,102323, and HD\,165014 are
accompanied by closely located sources in the line of sight, but
they are clearly separated in the AKARI 18\,$\mu$m image.  
Therefore, we conclude that spurious detections are not among our debris-disk candidates. 

\begin{figure*}[!t]
\includegraphics[width=7cm]{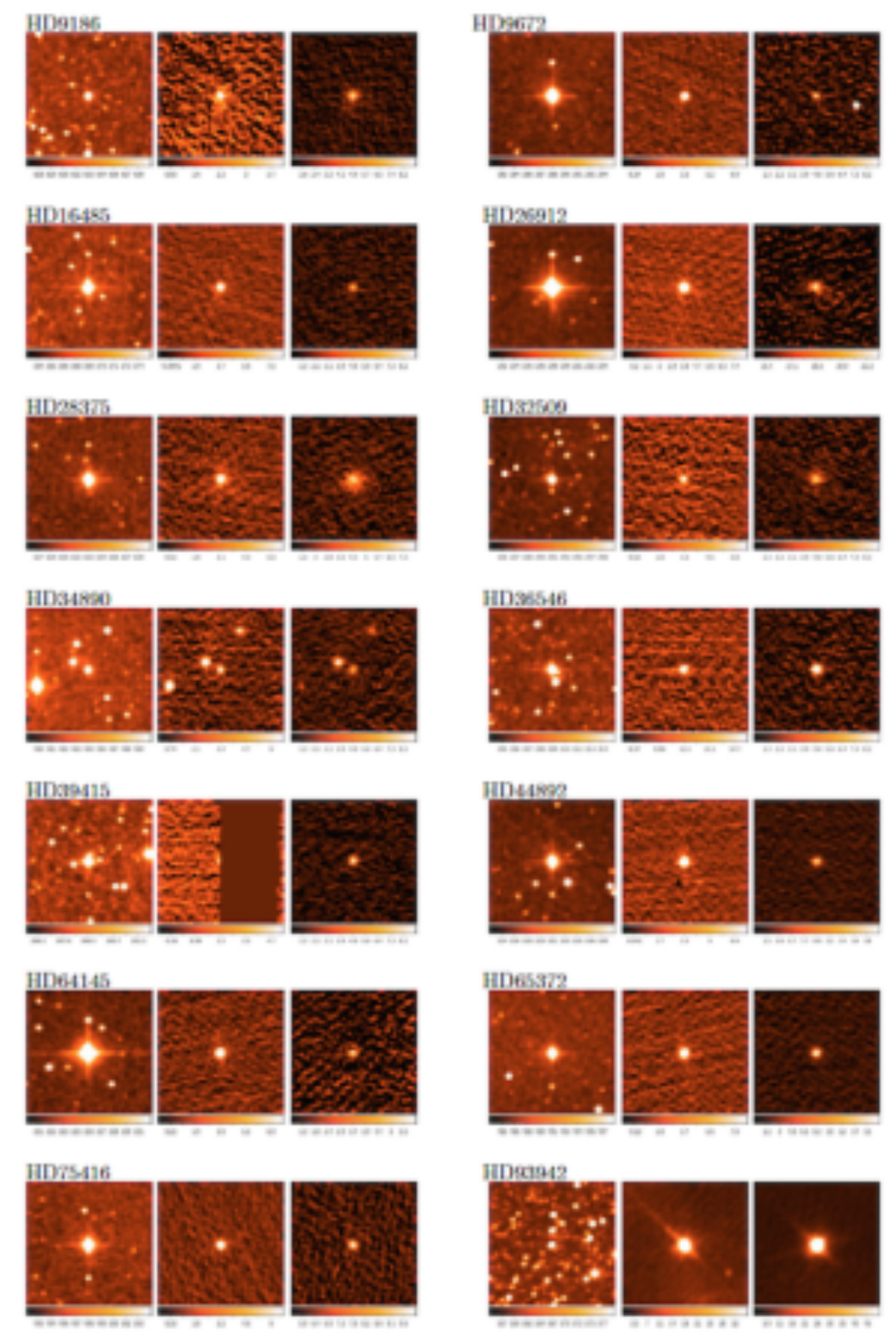}
\includegraphics[width=7cm]{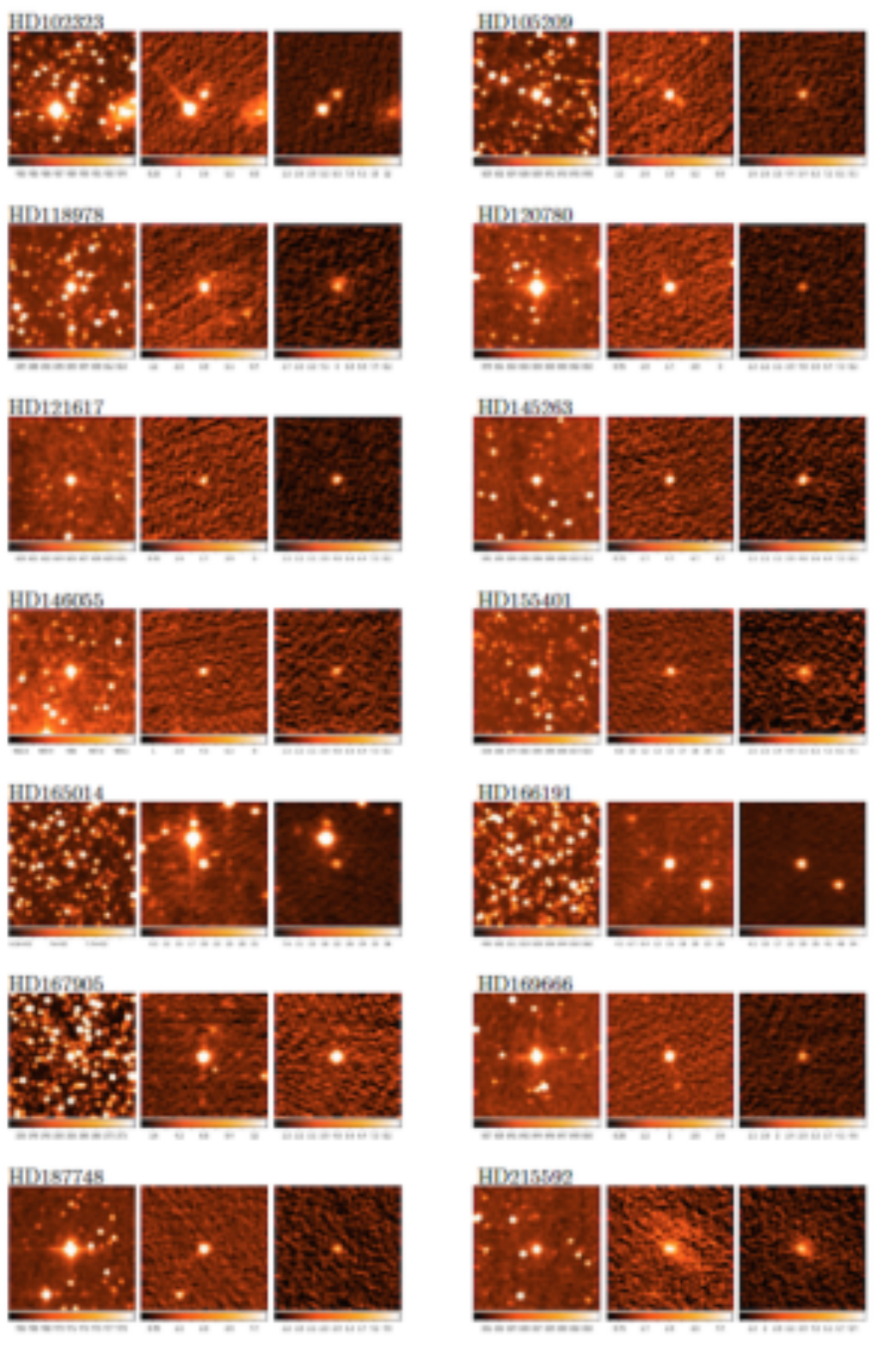}
\caption{2MASS Ks, AKARI\,9\,$\mu$m, and AKARI\,18$\mu$m, 
images (from left to right) for 53 debris disks candidates.}
\label{fig:cross-id}
\end{figure*}

\addtocounter{figure}{-1}
\begin{figure*}
\includegraphics[width=7cm]{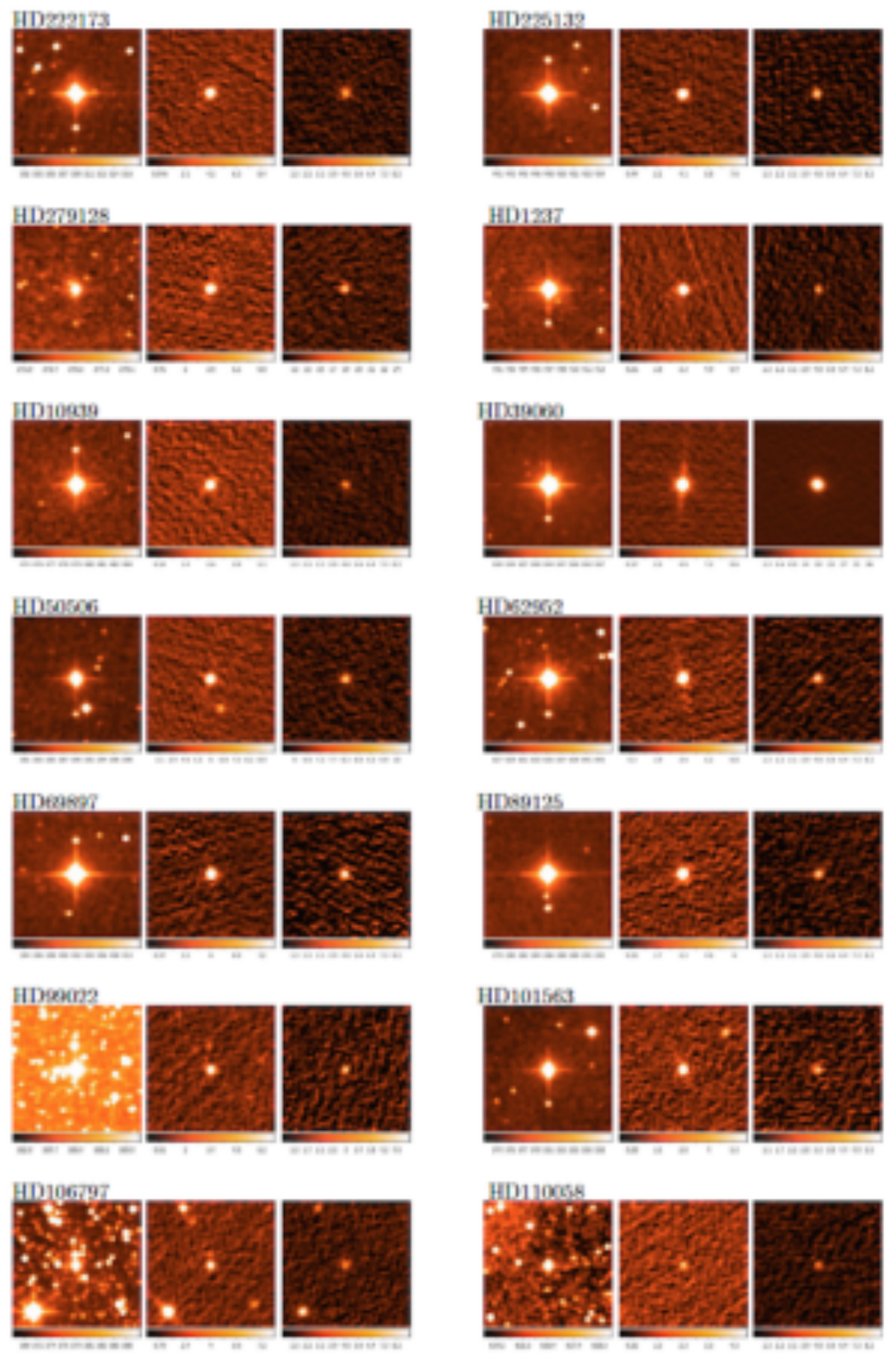}
\includegraphics[width=7cm]{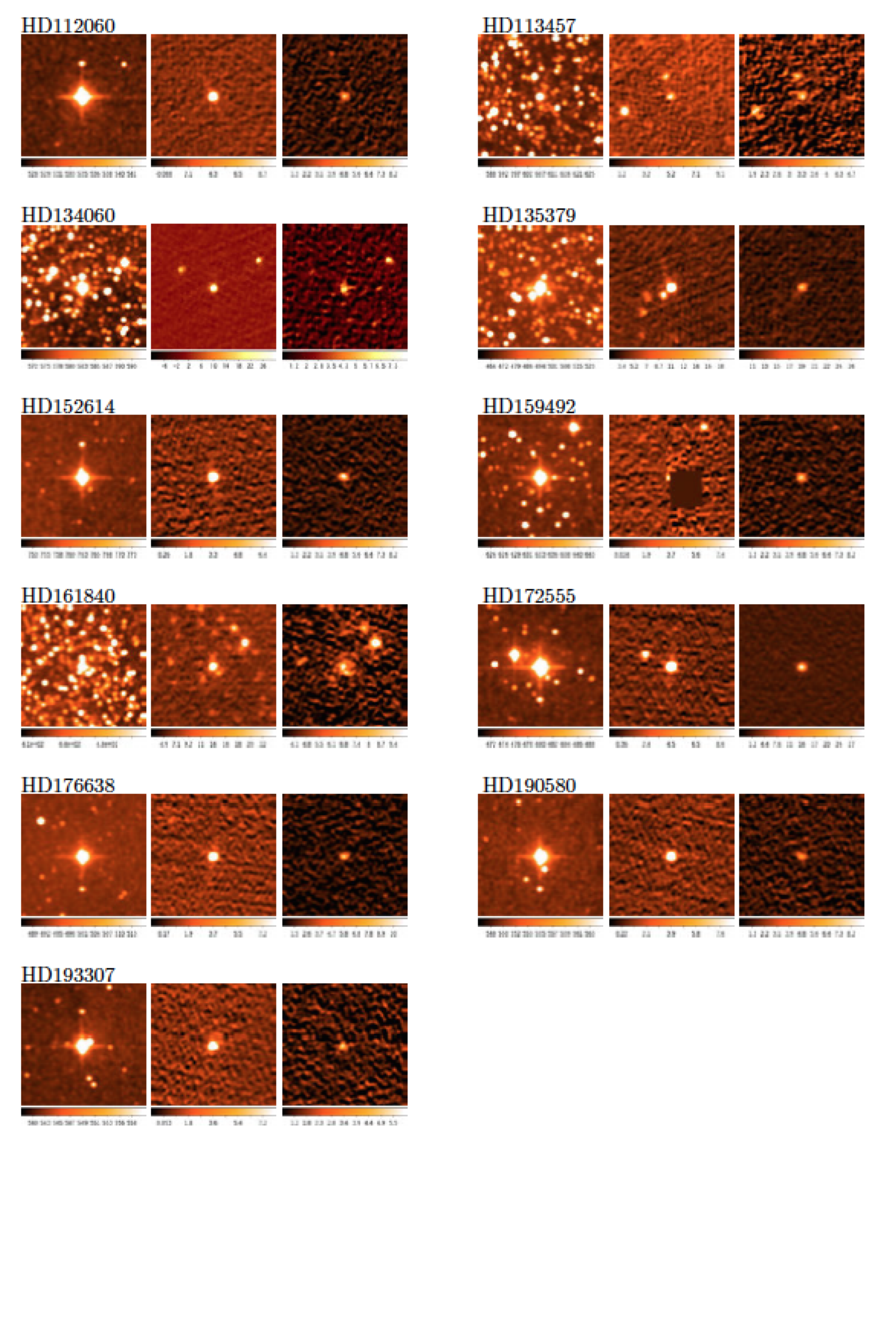}
\caption{Continued.}
\label{fig:ftex}
\end{figure*}


\section{Collisional evolution}
\label{appendix_col}

In a quasi steady-state collisional cascade, the flux-ratio evolution is
given by \citep[e.g.,][]{Kobayashi10}, 
\begin{equation}
 \frac{F_{\rm disk}}{F_{\ast}} = 
\frac{1}{1+t/\tau_0}
\frac{F_{\rm disk,0}}{F_{\ast}},\label{eq:flux_evo_ori} 
\end{equation}
where $F_{\rm disk,0}$ is the initial disk flux and $\tau_0$ is the
initial collisional cascade timescale, given by \citet{Kobayashi10} as,
\begin{eqnarray}
\tau_0 &\approx& 1.5 \left( \frac{s_{\rm p}}{\rm 3000\,km} \right)^{1.92}
\left( \frac{M_{\rm tot,0}}{M_{\oplus}} \right)^{-1}
\left( \frac{R}{\rm 2.5\,AU} \right)^{4.18} \nonumber\label{eq:tcol} \\
&&
\times
\left( \frac{\Delta R}{\rm 0.4\,R} \right)
\left( \frac{e}{\rm 0.1} \right)^{-1.4} \,\rm{Gyr},  
\end{eqnarray}
where $s_{\rm p}$ is the planetesimal radius, the radius of largest
bodies in the collisional cascade, $M_{\rm tot,0}$ is the initial total mass of
bodies, $M_\oplus$ is the mass of the Earth, $R$ is the radius of the planetesimal
belt, $\Delta R$ is the width of the belt, and $e$ is the eccentricity
of planetesimals. 
For the derivation of Eq.~(\ref{eq:flux_evo_ori}), a steady-state
collisional cascade is assumed. The steady state is achieved
at $t \gg \tau_0$ and the flux ratio depends on the initial condition
at $t \ll \tau_0$. Therefore we additionally assume $t \gg \tau_0$ and
then eq.~\ref{eq:flux_evo_ori} becomes
\begin{equation}
 \frac{F_{\rm disk}}{F_\ast} \approx \frac{\tau_0}{t} \frac{F_{\rm
  disk,0}}{F_\ast}. 
\end{equation}

In the steady-state collisional cascade, the surface number
density of bodies with radii from $s$ to $s+ds$, $n_{\rm s} (s) ds$, is
proportional to $s^{1-p}$, where $p$ is a constant.  The power-law index
$p$ is determined by the dependence of collisional velocity and
collisional strength on the radii of bodies \citep{Kobayashi10}. The
collisional strength is governed mainly by material properties for $s\la
1\,$km and by gravity $s \ga 1$\,km \citep[e.g.,][]{benz}. According to
\citet{Kobayashi10} based on the mass dependence of strength obtained by
hydrodynamic simulations \citep{benz}, we assume $p \approx 3.66$ for $s
< 1\,$km and $p \approx 3.04$ for $s > 1$\,km. 
The radius of smallest bodies is set to be 1\,$\mu$m.
For blackbody dust, this size distribution gives
\begin{eqnarray}
 \frac{F_{\rm disk,0}}{F_{\ast}} &\approx& 0.91 
  \left(\frac{M_{\rm tot,0}}{M_\oplus}\right) 
  \left(\frac{s_{\rm p}}{3000\,{\rm km}}\right)^{-0.96}
  \nonumber \\
  && \times 
\left(\frac{B_\nu(T_{\rm d})/B_{\nu}(T_\ast)}{1.6\times 10^{-3}}\right), 
\label{eq:ratio} 
\end{eqnarray}
where $B_\nu$ is the Planck function, $T_\ast$ and $T_{\rm d}$ are,
the stellar and dust temperatures, respectively, 
the value of
$B_\nu (T_\ast) / B_\nu (T_{\rm d})$ is estimated 
for $T_{\rm d} = 180$\,K, $T_* = 5,800$\,K,
and $\nu$ for the wavelength of 18\,$\mu$m, 
and we assume that
$s_{\rm p} \gg 1$\,km for this derivation.  From
Eqs.~(\ref{eq:tcol})--(\ref{eq:ratio}), we obtain Eq.~(\ref{eq:t0}).
As shown in Eq.~(\ref{eq:t0}), $t_0$ is independent of the total mass
of bodies and the width of the planetesimal belt. 

\end{appendix}
\end{document}